\documentclass[aps,twocolumn,superscriptaddress,showpacs,showkeys]{revtex4}
\usepackage{graphics,graphicx,dcolumn,bm,fleqn,epic,eepic,float}
\usepackage{amssymb,amsmath,multirow,rotate,color,float,times}
\usepackage{color}
\usepackage{soul}                    
\definecolor{red}{rgb}{1,0,0}
\definecolor{green}{rgb}{0,0.5,0}
\definecolor{blue}{rgb}{0,0,1}

\newcommand{\p}{\partial}

\newcommand{\fracpp}[1]{\frac{\p}{\p#1}}
\newcommand{\eps}{\varepsilon}

\newcommand{\veclambda}{{\boldsymbol{\lambda}}}
\newcommand{\vecalpha}{{\boldsymbol{\alpha}}}
\newcommand{\vecbeta}{{\boldsymbol{\beta}}}
\newcommand{\vecgamma}{{\boldsymbol{\gamma}}}

\newcommand{\vecDDrift}{\mathbf{D}^{(1)}}
\newcommand{\vecDDiff}{\mathbf{D}^{(2)}}
\newcommand{\e}[1]{{ e^{#1/\theta} }}
\begin{document}

\title{Analyzing a stochastic process driven by Ornstein-Uhlenbeck noise}

\author{B.~Lehle}
\author{J.~Peinke}
\affiliation{Institute of Physics, University of Oldenburg, D-26111 Oldenburg, Germany}

\begin{abstract}
A scalar Langevin-type process $X(t)$ that is driven by Ornstein-Uhlenbeck noise $\eta(t)$ is non-Markovian.
However, the joint dynamics of $X$ and $\eta$ is described by a Markov process in two dimensions.
But even though there exists a variety of techniques for the analysis of Markov processes, it is still a challenge to
estimate the process parameters solely based on a given time series of $X$.
Such a partially observed 2D-process could, e.g., be analyzed in a Bayesian framework using Markov chain Monte Carlo methods.
Alternatively, an embedding strategy can be applied, where first the joint dynamic of $X$ and its temporal derivative
$\dot X$ is analyzed. Subsequently the results can be used to determine the process parameters of $X$ and $\eta$.
In this paper, we propose a more direct approach that is purely based on the moments of the increments of $X$, which
can be estimated for different time-increments $\tau$ from a given time series.
From a stochastic Taylor-expansion of $X$, analytic expressions for these moments can be derived, which can be used to
estimate the process parameters by a regression strategy.
\end{abstract}




\maketitle

\section{Introduction}
\label{sec_intro}

\noindent A stochastically forced, first order differential equation provides an appropriate description for the evolution
of many physical, chemical or biological systems. For simplicity, we restrict ourselves to the evolution of the
scalar quantity $X(t)$ in the following. Additionally, we assume that the coefficient functions in its evolution equation
do not explicitely depend on time. Thus, we consider an equation of  the form
\begin{eqnarray}\label{evolution_X}
\fracpp{t} X &=& f(X) +g(X)\,\eta(t),
\end{eqnarray}

\noindent where $\eta(t)$ denotes the stochastic force. Such an equation arises not only for the
"obvious" case, where a deterministic system is driven by some external stochastic force, but also for complex dynamical
systems consisting of a large number of subsystems. Here, the phenomenon of self-organisation can give rise to a dynamic
of so-called order parameters that "enslave" the dynamic of the microscopic subsystems
\cite{haken04}, leading to an equation of the above form. However, now the stochastical force $\eta(t)$ can
no longer be considered to be external but is an intrinsic part of the system dynamic.

So far, the statistical properties of $\eta(t)$ have not been specified. In practice, this force quite often is treated as
Gaussian white noise. Frequently, the central limit theorem can be invoked, which then justifies the assumption of a Gaussian
probability density. The assumption of delta-correlated noise, however, is an idealization. Real-world systems usually have
some finite correlation time $\theta$.
How strong the correlations of $\eta(t)$ affect the statistics of $X(t)$ depends on the ratio of $\theta$ and the
characteristic time-scale $T$ of $X(t)$ \cite{haenggi95}. For $\theta\ll T$ the force $\eta(t)$ can be approximated by
delta-correlated noise, leading to a Markovian description. The probably most famous example is given by Einsteins description
of Brownian motion by a Wiener-process \cite{einstein05}. Even though the true process is non-Markovian on a microscopic scale,
the Markov property can be taken for given for increments larger than some limit timescale. This approach has also successfully
been applied to other problems like the description of turbulent velocity increments by a process in scale
\cite{friedrich97,renner01}. Here, the limit timescale is replaced by its spatial analogon, which is
denoted as Markov-Einstein coherence length in \cite{lueck06}.

Although there are many systems where $\theta$ is sufficiently small and can be neglected, in a variety of systems such an
idealization leads to notable differences.
For such systems, it is no longer justified to ignore the correlations of $\eta(t)$.
However, if we want to account for these correlations, we need a description of the evolution of $\eta(t)$ that goes beyond
a "purely random Gaussian process" \cite{risken89}.
The most natural and simple generalization of Gaussian white noise is provided by exponentially correlated Gaussian
noise, as generated by a stationary Ornstein-Uhlenbeck process.
Even if this is not the most general description of colored noise, the assumption that $\eta(t)$ is a stationary process
obeying
\begin{eqnarray}\label{evolution_eta}
\fracpp{t} \eta &=& -\frac{1}{\theta} \eta+\frac{1}{\theta}\xi(t),\qquad \theta>0,
\end{eqnarray}

\noindent makes Eq.~(\ref{evolution_X}) applicable to a much larger class of problems.
Here, $\xi(t)$ denotes Gaussian white noise with $\left<\xi(t)\right>\!=\!0$ and
$\left<\xi(t)\xi(t')\right>\!=\!\delta(t\!-\!t')$.
The characteristic time-scale of $\eta(t)$ is determined by the parameter $\theta$.
In the limit $\theta\to 0$, the case of Gaussian white noise is recovered.

Based on this assumption for $\eta(t)$, Eqs.~(\ref{evolution_X}) and (\ref{evolution_eta}) describe a Markov process in
two dimensions. However, the analysis of this process is hampered by the fact that usually only a 1D-series of values
of $X(t)$ will be available in practice.
In the mathematical community this problem is known as "partially observed diffusions" \cite{dembo86,campillo89}.
There are approaches to deal with such problems. In a Bayesian framework, e.g., one could use Markov chain
Monte Carlo methods for an estimation of $f$, $g$ and $\theta$ (see, e.g., \cite{golightly08}).
Alternatively, an embedding approach could be used, where first a series of velocities $\dot X(t)$ is calculated from the
values of $X(t)$ and subsequently the 2D-system $[X(t),\dot X(t)]$ is analyzed.
The drift- and diffusion functions of this 2D-system can then be used to determine $f$, $g$ and $\theta$. However, some care
has to be taken with this latter approach, because the velocities need to be estimated numerically. This leads to spurious
correlations that may affect the results \cite{lehle15}.

Here, we propose a more direct approach that is purely based on the moments of the conditional increments
\begin{eqnarray}
\Delta X(\tau,t_0)\big|_{x_0} &:=& X(t_0+\tau)\big|_{x_0}\!-x_0,
\end{eqnarray}

\noindent where $(..)|_{x_0}$ denotes conditioning on $X(t_0)=x_0$.
In the following, we restrict ourselves to a statistically stationary process $X(t)$, i.e., the moments of
$\Delta X$ do not depend on $t_0$ and can be estimated from a single time series of $X(t)$ for different values
of $\tau$ and $x_0$.
Our strategy for parameter estimation is based on a stochastic Taylor expansion, which allows us to express the increments
$\Delta X$ by an infinite sum that involves multiple integrals with respect to a noise generating process.
Since the Ornstein-Uhlenbeck process can be solved analytically, explicit expressions for the correlations of these integrals
can be given.
Assuming $\tau$ and $\theta$ to be small as compared to the characteristical time-scale of $X(t)$, the moments of $\Delta X$
can be approximated by a finite base of functions $r_i(\tau,\theta)$ that are weighted by coefficients
$\lambda_i(x_0,\theta)$.
These functions are used to fit the moments of $\Delta X$ and thus allow for an estimation of the process parameters.
In a first step, the parameter $\theta$ is estimated by a non-linear minimization procedure. Subsequently, by
linear fitting, the coefficients $\lambda_i$ are estimated, which are then used to determine $f$ and $g$.

It may be noted that in the limit $\theta\to 0$ the functions $r_i$ reduce to powers of $\tau$. The above approach
then recovers the so-called direct estimation method, which is applicable to processes driven by Gaussian white
noise. Due to its simplicity of use, this method has found wide-spread use. For an overview see e.g. \cite{friedrich11}.

The standard rules for integro-differential equations apply to the calculations in this paper for $\theta>0$, i.e., for
correlated driving noise.
Therefore, we adopt the Stratonovich definition of stochastic integrals in the limit $\theta=0$.
By this choice, the results will remain valid in unchanged form also in the white-noise limit \cite{vankampen81}.

This paper is structured as follows. In Sec.~\ref{sec_noise}, we compile some properties of the stochastic force
$\eta(t)$. Subsequently, the stochastic Taylor expansion of $X(t)$ is given in Sec.~\ref{sec_Taylor},
which is used in Sec.~\ref{sec_moments} to provide a series representation for the moments of the conditional
increments of $X$. The functional form of the series terms is discussed in Sec.~\ref{sec_expectation_J}. Subsequently,
a series truncation is performed in Sec.~\ref{sec_approximate}, which then is used in Sec.~\ref{sec_estimation}
to formulate a strategy for parameter estimation. To verify the analytical
results, a numerical example will finally be given in Sec.~\ref{sec_example}.

\section{Stochastic force}
\label{sec_noise}

\noindent We assume that $\eta(t)$ is a stationary Ornstein-Uhlenbeck process obeying Eq.~(\ref{evolution_eta}).
Ornstein-Uhlenbeck processes are well understood and can be solved analytically.
The realization of $\eta(t\ge t_0)$ can explicitely be expressed in terms of an initial value $\eta(t_0)$ and the
realization of $\xi(t\ge t_0)$.
Since we focus in the following on the stationary process, we are free to choose $t_0\equiv 0$, which simplifies notation.
One then finds
\begin{eqnarray}\label{def_eta}
\eta(t) &=& \eta(0)\,e^{-t/\theta}+\frac{1}{\theta} \int_{0}^t e^{(s-t)/\theta}\xi(s)\,ds.
\end{eqnarray}

\noindent This equation holds for arbitrary values of $\eta(0)$. It describes a realization of the process $\eta(t)$,
i.e., a trajectory in time, in terms of $\eta(0)$ and a trajectory of $\xi(t)$.
Expectation values of functionals of $\eta(t)$ thus are obtained by averaging over the realizations of $\eta(0)$ and $\xi(t)$.
As these quantities are statistically independent, averaging may be performed in two steps using
\begin{eqnarray}
\big<..\big>_{\xi,\eta(0)} &=& \big<\big<..\big>_\xi\big>_{\eta(0)} = \big<\big<..\big>_{\eta(0)}\big>_\xi.
\end{eqnarray}

\noindent In the next section, $\eta(t)$ will be expressed as derivative of a noise-generating process $V(t)$ with
$V(0)\equiv 0$. This implies
\begin{eqnarray}\label{def_V}
V(t) &:=& \int_0^t \eta(s)\, ds.
\end{eqnarray}

\noindent This process plays a comparable role for $\eta(t)$ as the Wiener process does for Gaussian white noise.
Using Eq.~(\ref{def_eta}), we also may describe a trajectory of $V(t)$ directly in terms of $\eta(0)$ and a trajectory
of $\xi(t)$,
\begin{eqnarray}\label{def2_V}
V(t) &=& \eta(0)\,\theta(1-e^{-t/\theta}) \cr
     &&  +\int_0^t\left[1-e^{(s-t)/\theta}\right]\!\xi(s)\,ds.
\end{eqnarray}

\noindent It may easily be checked that $V(t)$ approaches the Wiener process $W(t):=\int_0^t \xi(s)\,ds$ in the limit
$\theta\to 0$.

Finally, we consider some properties of the stationary process. With Eq.~(\ref{def_eta}) the
stationary probability density function (PDF) of $\eta(t)$ is found to be a Gaussian with variance $1/(2\theta)$ and
vanishing mean. Furthermore, the autocorrelation function is found to be
\begin{eqnarray}
\left<\eta(t)\eta(t')\right> &=& \frac{1}{2\theta}e^{-|t-t'|/\theta}.
\end{eqnarray}

\noindent This means that $\eta(t)$ is normalized in the following sense: its strength, i.e., the integral over its
autocorrelation function, is constant and equals unity.
Therefore, in the limit $\theta\!\to\!0$ the autocorrelation approaches $\delta(t-t')$ and $\eta(t)$ approaches
Gaussian white noise $\xi(t)$.

\section{Stochastic Taylor expansion}
\label{sec_Taylor}

\noindent Since, later on, we will focus on moments of conditional increments of $X(t)$, we need an analytic description
of these increments. Assuming smooth functions $f$ and $g$, such a description can be provided by a stochastic Taylor
expansion, which allows us to express a trajectory of $X(t)$ in terms of $X(0)$, values and derivatives of $f$ and $g$
at $X(0)$, and a trajectory of $\eta(t)$. Such expansions are described in great detail, e.g., in \cite{platen99}, and
we will closely follow these lines.

The starting point is Eq.~(\ref{evolution_X}) in the form $dX=fdt+g\eta dt$. Expressing $\eta(t)$ as derivative of
a noise generating process $V(t)$ as defined by Eq.~(\ref{def_V}), this may be written as
\begin{eqnarray}\label{def_dX}
dX(t) &=& f[X(t)]\,dt+g[X(t)]\,dV(t).
\end{eqnarray}

\noindent Next, the infinitesimal increment of an arbitrary, smooth, function $h(X)$ is considered. Since $X(t)$ is
continuously differentiable, the standard chain-rule of differentiation applies,
\begin{eqnarray}
dh[X(t)] &=& \frac{\p h[X(t)]}{\p X(t)} \,dX(t).
\end{eqnarray}

\noindent Expressing $dX$ by Eq.~(\ref{def_dX}) and introducing the operators
\begin{subequations}\label{def_L0L1}
\begin{eqnarray}
L_0 &:=& f[X(t)]\frac{\p}{\p X(t)},\\
L_1 &:=& g[X(t)]\frac{\p}{\p X(t)},
\end{eqnarray}
\end{subequations}

\noindent this may be written as
\begin{eqnarray}\label{ottos_lemma}
dh\{t\} &=& [L_0 h]\{t\}\,dt +[L_1 h]\{t\}\,dV(t).
\end{eqnarray}

\noindent Here, we use the notation $\{t\}$ to indicate that all arguments of a function or expression are to
be evaluated at time $t$. Now the actual expansion of $h$ can be started. In integral form, Eq.~(\ref{ottos_lemma}) reads
\begin{eqnarray}\label{otto_expansion_first}
h\{t\} &=& h\{0\}+\int_0^t [L_0 h]\{s\}\,ds\cr
       && +\int_0^t[L_1 h]\{s\}\,dV(s).
\end{eqnarray}

\noindent Since we considered $f$, $g$ and $h$ to be smooth, $L_0 h$ and $L_1 h$ are also smooth functions of $X$.
Consequently, Eq.~(\ref{otto_expansion_first}) can be applied,
\begin{eqnarray}
[L_0 h]\{s\} &=& [L_0 h]\{0\}+\int_0^s [L_0L_0 h]\{s'\}\,ds'\cr
             && +\int_0^s[L_1L_0 h]\{s'\}\,dV(s'),\\~
[L_1 h]\{s\} &=& [L_1 h]\{0\}+\int_0^s [L_0L_1 h]\{s'\}\,ds'\cr
             && +\int_0^s[L_1L_1 h]\{s'\}\,dV(s').
\end{eqnarray}

\noindent Inserting these results into Eq.~(\ref{otto_expansion_first}) then yields
\begin{eqnarray}\label{otto_expansion_second}
h\{t\} &=& h\{0\}
+\int_0^t[L_0 h]\{0\}\,ds\cr
&& +\int_0^t\int_0^s [L_0L_0 h]\{s'\}\,ds'\,ds\cr
&& +\int_0^t\int_0^s[L_1L_0 h]\{s'\}\,dV(s')\,ds\cr
&& +\int_0^t[L_1 h]\{0\}\,dV(s)\cr
&& +\int_0^t\int_0^s [L_0L_1 h]\{s'\}\,ds'\,dV(s)\cr
&& +\int_0^t\int_0^s[L_1L_1 h]\{s'\}\,dV(s')\,dV(s) .
\end{eqnarray}

\noindent The single integrals from Eq.~(\ref{otto_expansion_first}), which had time-dependent integrands
$[L_ih]\{s\}$, are now replaced by single integrals with constant integrands $[L_ih]\{0\}$ plus additional double integrals
with time-dependent integrands $[L_iL_jh]\{s'\}$. Expressing these functions by Eq.~(\ref{otto_expansion_first}) will put
the game on the next level, leading to constant double integrals plus variable triple integrals --- and so on. In
the end, one is left with an infinite sum of multiple integrals, which only depend on $t$ and the realization of $V(t)$,
that are multiplied by coefficient functions that only depend on values and derivatives of $f$, $g$ and $h$ at $X(0)$,
\begin{eqnarray}
h\{t\} &=& h\{0\} +[L_0 h]\{0\}\int_0^t\!ds\cr
&& +[L_1 h]\{0\}\int_0^t\!dV(s)\cr
&& +[L_0L_0 h]\{0\}\int_0^t\!\int_0^s\!ds'\,ds \cr
&& +[L_1L_0 h]\{0\}\int_0^t\!\int_0^s\!dV(s')\,ds\cr
&& +\ldots
\end{eqnarray}

\noindent Using a multi-index $\vecalpha$, defined as
\begin{eqnarray}
\vecalpha &:=& (\alpha_1,\ldots,\alpha_n),\quad n\in{\mathbb N},\quad
\alpha_i \in \{0,1\},
\end{eqnarray}

\noindent the expansion of $h[X(t)]$ can be compactly written as
\begin{eqnarray}
h[X(t)] &=& h[X(0)]+\sum_\vecalpha c_\vecalpha[X(0)] J_\vecalpha(t),
\end{eqnarray}

\noindent where the coefficient functions $c_\vecalpha$ are given by
\begin{eqnarray}
c_{(\alpha_1,\ldots,\alpha_n)}[X(0)] &:=& [L_{\alpha_1}\ldots L_{\alpha_n} h]\{0\},
\end{eqnarray}

\noindent and the integrals $J_\vecalpha$ by
\begin{eqnarray}\label{def_J_alpha}
J_{(\alpha_1,\ldots,\alpha_n)}(t) &:=&
\int_{s_n=0}^{t}
\int_{s_{n\!-\!1}=0}^{s_n}\!\cdots
\int_{s_1=0}^{s_2} \nonumber\\[.3em]
&&  \times \,dZ_{\alpha_1}\!(s_1)\cdots dZ_{\alpha_n}\!(s_n) ,
\end{eqnarray}
\noindent with
\begin{eqnarray}
dZ_j(s) &:=& \left\{\begin{array}{ll}
ds    &\; , j=0 \\
dV(s) &\; , j=1
\end{array}\right. .
\end{eqnarray}

\noindent In general, these integrals are functionals of the realization of $V(t)$ respectively $\eta(t)$ and thus stochastic
quantities. Only for $\alpha_1=\ldots=\alpha_n=0$ the integrals become purely deterministic and evaluate to
\begin{eqnarray}\label{J_deterministic}
J_{(0,\ldots,0)}(t) &=& \frac{1}{n!}\,t^n.
\end{eqnarray}

\noindent So far, the expansion of some arbitrary function $h(X)$ has been considered. Being interested in the
expansion of $X(t)$ itself, we choose $h(X)\equiv X$ in the following. Additionally, we fix the value $X(0)$ to
$x_0$, which then leaves us with
\begin{eqnarray}\label{expansion_X}
X(t)\big|_{x_0} &=& x_0 +\sum_\vecalpha c_\vecalpha(x_0)\, J_\vecalpha(t)\big|_{x_0},
\end{eqnarray}

\noindent where the coefficient functions are now defined as
\begin{eqnarray}\label{def_c_alpha}
c_{(\alpha_1,\ldots,\alpha_n)}(x_0) &:=& [L_{\alpha_1}\ldots L_{\alpha_n} X]\{0\}\big|_{x_0}.
\end{eqnarray}

\noindent Omitting arguments and using a prime to denote derivatives with respect to  $X$, the first few of
these functions (to be evaluated at $x_0$) read
\begin{eqnarray}
c_{(0)} &=&  f,\quad c_{(0,0)} = ff',\quad c_{(1,0)} = gf',\quad \ldots \cr
c_{(1)} &=&  g,\quad c_{(0,1)} = fg',\quad c_{(1,1)} = gg',\quad \ldots \;.
\end{eqnarray}

\noindent The conditioning of $J_\vecalpha$ in Eq.~(\ref{expansion_X}) deserves some comment. After all, a realization
of $J_\vecalpha(t)$ does not depend on $X(0)$ but is a pure functional of $\eta(t)$, which itself is a functional of
$\eta(0)$ and the realization of $\xi(t)$.
However, the PDF of $\eta(0)$ will, in general, depend on $X(0)$ [see, e.g., Appendix \ref{app_expectation_eta0} for the
expectation value of $\eta(0)|_{x_0}$].
As a consequence, ensemble averages of any conditioned functional $F[\eta(t)]|_{x_0}$ need to be calculated by averaging
over the realizations of $\xi(t)$ and over the {\em conditional} realizations $\eta(0)|_{x_0}$. Averaging may still be
performed in two steps, e.g., by
\begin{eqnarray}\label{ave_func_eta_x}
\Big<F\big[\eta(t)\big]\big|_{x_0}\Big> &=& \Big<\Big<F\big[\eta(t)\big]\Big>_\xi\Big>_{\eta(0)|_{x_0}}.
\end{eqnarray}

\section{Conditional moments of $\Delta X$}
\label{sec_moments}

\noindent We now turn to mean and variance of the conditional process increments of $X(t)$,
\begin{subequations}
\begin{eqnarray}
M^{(1)}(\tau,x) &:=& \Big<\Delta X(\tau)\big|_x\Big>,\\
M^{(2)}(\tau,x) &:=& \Big<\big[\Delta X(\tau)\big|_x-M^{(1)}(\tau,x)\big]^2\Big>,
\end{eqnarray}
\end{subequations}
\noindent where the increments are denoted by
\begin{eqnarray}
\Delta X(\tau)\big|_x &:=& X(t+\tau)\big|_{X(t)=x}-x.
\end{eqnarray}

\noindent Since we are conditioning on the value of $X$ at some arbitrary time $t$, we denote this value by $x$ instead of
by $x_0$. Additionally, we suppress the function argument $t$, because the statistical properties of the increments $\Delta X$ do
not depend on time for a stationary process. Stationarity also implies that the moments $M^{(k)}$ can be estimated from a
given time series of $X(t)$ by replacing the above ensemble-averages by time-averages (tacitly assuming ergodicity).
Using the results from the previous section, we already have an analytical description for the increments,
\begin{eqnarray}
\Delta X(\tau)\big|_x &=& \sum_\vecalpha c_\vecalpha(x) J_\vecalpha(\tau)\big|_x.
\end{eqnarray}

\noindent Hence, the conditional moments are given by
\begin{subequations}\label{series_Mk}
\begin{eqnarray}\label{series_M1}
M^{(1)}(\tau,x) &=& \sum_\vecalpha c_\vecalpha(x) \phi_\vecalpha(\tau,x),\\\label{series_M2}
M^{(2)}(\tau,x) &=& \sum_{\vecalpha,\vecbeta} c_\vecalpha(x) c_\vecbeta(x) \phi_{\vecalpha,\vecbeta}(\tau,x),
\end{eqnarray}
\end{subequations}
\noindent with (omitting arguments)
\begin{subequations}
\begin{eqnarray}
\phi_\vecalpha &:=& \big<J_\vecalpha\big|_x\big>,\\
\phi_{\vecalpha,\vecbeta} &:=& \big<J_\vecalpha\big|_x J_\vecbeta\big|_x\big>
                                      -\big<J_\vecalpha\big|_x\big>\big<J_\vecbeta\big|_x\big>.
\end{eqnarray}
\end{subequations}

\noindent As a result, we now have analytic descriptions of the moments --- but unfortunately in terms of infinite series.
In order to obtain approximate descriptions with a finite number of terms, the functional form of $\phi_\vecalpha(\tau,x)$
needs to be investigated in the following. This also provides us with the functional form of
$\phi_{\vecalpha,\vecbeta}(\tau,x)$, because a product $J_\vecalpha J_\vecbeta$ can be expressed by a sum of integrals
$J_\vecgamma$ (see Appendix \ref{app_products_j_integrals}),
\begin{eqnarray}
J_\vecalpha J_\vecbeta &=& \sum_{\vecgamma\in{\cal M}(\vecalpha,\vecbeta)} J_\vecgamma,
\end{eqnarray}
\noindent which implies
\begin{eqnarray}\label{phi_ab_by_phi_a}
\phi_{\vecalpha,\vecbeta} &=& \sum_{\vecgamma}\phi_\vecgamma -\phi_\vecalpha\phi_\vecbeta.
\end{eqnarray}

\section{Functional form of $\phi_\vecalpha$}
\label{sec_expectation_J}

\noindent The starting point for the calculation of $\phi_\vecalpha$ is the definition of the integral $J_\vecalpha$, as
provided by Eq.~(\ref{def_J_alpha}). Let us consider an index vector $\vecalpha$ of length $n$ and denote the
number of its non-zero entries by $m$. Using $dV=\eta\,dt$, Eq.~(\ref{def_J_alpha}) may then be written as $n$-fold
integral with respect to time over an $m$-fold product of $\eta$,
\begin{eqnarray}
J_{\vecalpha}(\tau) &=&
\int_{\Omega(\tau)} \Big[\prod_{\alpha_j=1}\eta(s_j)\Big] \,ds_1\cdots ds_n.
\end{eqnarray}

\noindent Here, the shortcut $\Omega(\tau)$ has been introduced to denote the integration domain (a simplex
in ${\mathbb R}^n$ with $0\le s_i\le s_{i+1}$ and $s_n\le\tau$).
The ensemble average of $J_\vecalpha\big|_x$ then reads
\begin{eqnarray}\label{def_phi_alpha_explicit}
\phi_{\vecalpha}(\tau,x) = \int_{\Omega(\tau)}\!\!\! C_\eta(s_{j_1},\ldots,s_{j_m},x) \,ds_1\cdots ds_n,
\end{eqnarray}

\noindent where the values $j_1,\ldots,j_m$ denote the positions of the non-zero entries in $\vecalpha$ and
$C_\eta$ the $m$-point correlation function of $\eta(t)\big|_x$. For arbitrary times $t_1,\ldots,t_m$ this function is
defined as
\begin{eqnarray}\label{def_C_eta}
C_\eta(t_1,\ldots,t_m,x) :=
\big<\big<\eta(t_1)\cdots\eta(t_m)\big>_\xi\big>_{\eta(0)|_x}.
\end{eqnarray}

\noindent According to Eq.~(\ref{def_eta}), $\eta$ may be splitted up into one part depending only on $\eta(0)$ and
another one depending only on $\xi$,
\begin{eqnarray}\label{split_eta}
\eta(t) &=& Y\,\frac{e^{-t/\theta}}{\sqrt{\theta}}+u(t),
\end{eqnarray}
\noindent with
\begin{eqnarray}\label{def_Y}
Y &:=& \sqrt{\theta}\,\eta(0),\\
\label{def_u}
u(t) &:=& \frac{1}{\theta} \int_{0}^t e^{(s-t)/\theta}\xi(s)\,ds.
\end{eqnarray}

\noindent Consequently, $C_\eta$ can be expressed in terms of conditional moments of
$Y$ and correlation functions of $u(t)$, denoted as
\begin{eqnarray}\label{def_C}
C(t_1,\ldots,t_k) &:=& \big<u(t_1)\cdots u(t_k)\big>_\xi.
\end{eqnarray}

\noindent For example, we find
\begin{eqnarray}
C_\eta(t_1,t_2,x) \!&=&
\big<Y^2|x\big>\frac{e^{-(t_1+t_2)/\theta}}{\theta}\nonumber\\
   \!&&+\big<Y|x\big>\frac{e^{-t_1/\theta}}{\sqrt{\theta}}C(t_2)\nonumber\\
   \!&&+\big<Y|x\big>\frac{e^{-t_2/\theta}}{\sqrt{\theta}}C(t_1)\nonumber\\
   \!&&+C(t_1,t_2).
\end{eqnarray}

\noindent Explicit expressions for $C$ can be found by virtue of Eq.~(\ref{def_u}). It turns out that the correlation
functions of $u(t)$ have the same structure as those of Gaussian white noise $\xi(t)$ (see Appendix~\ref{app_correlations_u}).
The $k$-point correlation of $u$ vanishes for odd values of $k$, while for even values it can be expressed by a sum
of products of the two-point correlation [Eq.~(\ref{twopoint_eta})].

Since $C(t_1,\ldots,t_k)$ vanishes for odd $k$, the expressions for $C_\eta$ may contain either only even or only odd
moments of $Y\big|_x$. To provide an example:
\begin{eqnarray}
C_\eta(t_1,t_2,t_3,x) \!&=&
\big<Y^3|x\big>\frac{e^{-(t_1+t_2+t_3)/\theta}}{\theta}\cr
   \!&&+\big<Y|x\big>\frac{e^{-t_1/\theta}}{\sqrt{\theta}}C(t_2,t_3)\cr
   \!&&+\big<Y|x\big>\frac{e^{-t_2/\theta}}{\sqrt{\theta}}C(t_1,t_3)\cr
   \!&&+\big<Y|x\big>\frac{e^{-t_3/\theta}}{\sqrt{\theta}}C(t_1,t_2).
\end{eqnarray}

\noindent With the above results, the intgral on the right-hand side of Eq.~(\ref{def_phi_alpha_explicit}) can be evaluated,
which leads to
\begin{eqnarray}
\phi_{\vecalpha}(\tau,x) = \sum_{k=0}^{2k\le m}\big<Y^{m-2k}|x\big>a_k(\tau).
\end{eqnarray}

\noindent Here, we introduced the shortcuts $a_k(\tau)$ to denote the functions that stem from the integrations with respect
to time. Actually, these functions depend also on the index vector $\vecalpha$, but we supressed this argument for notational
simplicity. For an explicit example see Appendix \ref{app_example_phialpha}.

Later on, it proves to be useful to re-arrange the right-hand side of this equation by expressing the
powers of $Y$ in terms of Hermite polynomials in $Y$. Reordering terms then yields 
\begin{eqnarray}
\phi_{\vecalpha}(\tau,x) = \sum_{k=0}^{2k\le m}\big<H_{m-2k}(Y)|x\big>b_k(\tau),
\end{eqnarray}

\noindent where the $k$-th Hermite polynomial is defined as
\begin{eqnarray}
H_k(y) := (-1)^ke^{y^2}\left(\frac{\p}{\p y}\right)^k\,e^{-y^2},
\end{eqnarray}

\noindent and the functions $b_k$ are linear combinations of the functions $a_k$. For example, a right-hand side of the
form $\big<Y^2|x\big>a_0+a_1$ becomes $\big<H_2(Y)|x\big>b_0+b_1$ with $b_0=a_0/4$ and $b_1=a_1+a_0/2$.

By mathematical induction, it may be shown that the functions $a_k(\tau)$, and thus also the functions $b_k(\tau)$, are
linear combinations of the functions
\begin{subequations}\label{r_ab_alpha}
\begin{eqnarray}
\tilde r_{0b}(\tau) &:=& \theta^{\ell(\vecalpha)}\big[1-e^{-b\tau/\theta}\big],\\
\tilde r_{a0}(\tau) &:=& \theta^{\ell(\vecalpha)}\left(\tau/\theta\right)^a,\\
\tilde r_{ab}(\tau) &:=& \theta^{\ell(\vecalpha)}\left(\tau/\theta\right)^a e^{-b\tau/\theta},
\end{eqnarray}
\end{subequations}
\noindent with
\begin{eqnarray}
a,b\in\mathbb{N},\quad a\le \ell(\vecalpha),\quad b\le m
\end{eqnarray}
\noindent and
\begin{eqnarray}
\ell(\vecalpha) &:=& n-\frac{m}{2} \;=\; \sum_{i=1}^n(1-\alpha_i/2).
\end{eqnarray}

\noindent We thus may express $b_k$ in the form $\sum\tilde \lambda_{ij}\tilde r_{ij}$. Note, that the coefficients
$\tilde \lambda_{ij}$ do not depend on $\theta$, because this dependency is completely accounted for
by the functions $\tilde r_{ij}$. This property will be useful in the next section, when we consider the magnitude of
individual terms.
However, since the functions $\tilde r_{ij}$ depend on $\ell(\vecalpha)$, this property can not be sustained for
the following base of functions $r_{ij}$, which is used to describe the $\tau$-dependency of $b_k$, and thus of
$\phi_\vecalpha$, for {\em arbitrary} vectors $\vecalpha$,
\begin{eqnarray}
{\cal B} &:=& \big\{r_{0b}(\tau)\big|b\in\mathbb{N}\big\}\cup
              \big\{r_{a0}(\tau)\big|a\in\mathbb{N}\big\}\cr
         && \cup\; \big\{r_{ab}(\tau)\big|a,b\in\mathbb{N}\big\},
\end{eqnarray}
\noindent with
\begin{subequations}\label{def_r_ab}
\begin{eqnarray}
r_{0b}(\tau) &:=& 1-e^{-b\tau/\theta},\\
r_{a0}(\tau) &:=& \frac{1}{a!}\,\tau^a,\\
r_{ab}(\tau) &:=& \left(\tau/\theta\right)^a e^{-b\tau/\theta}.
\end{eqnarray}
\end{subequations}

\noindent Note that the product of any two functions of this base lies in the linear span of $\cal B$. According
to Eq.~(\ref{phi_ab_by_phi_a}), therefore, $\cal B$ not only provides a base for the functions $\phi_\vecalpha$ but also
for the functions $\phi_{\vecalpha\vecbeta}$.

\section{Series truncation}
\label{sec_approximate}

\noindent With the results from the previous section, the series representation of the moments $M^{(k)}$,
Eq.~(\ref{series_Mk}), can now be expressed in terms of functions $r_i\in{\cal B}$,
\begin{eqnarray}\label{Mk_lambda_r}
M^{(k)}(\tau,x) &=& \sum_{r_i\in{\cal B}}\lambda^{(k)}_i(x)\,r_i(\tau),
\end{eqnarray}

\noindent where each coefficient function $\lambda_i$ consists of an infinite sum of terms. These terms, in general,
are formed by powers of $\theta$, the functions $c_\vecalpha$ from the Taylor expansion [Eq.~(\ref{def_c_alpha})],
and the expectation values $\left<H_n(Y)|x\right>$.

In order to approximate $M^{(k)}$ by a finite number of functions, it becomes necessary to make some assumptions on
the magnitude of the individual terms. First, we assume that $X(t)$ has been normalized to ensure a characteristic
time-scale of unity and coefficient functions $c_\alpha$ of order $O(1)$. Second, we assume
\begin{eqnarray}\label{magnitude_theta}
\tau = O(\eps),\qquad \theta = O(\eps^2),
\end{eqnarray}

\noindent where $\eps$ has been introduced to denote a quantity that is small as compared to unity. We may identify
$\eps$ as the largest increment $\tau$ for which we assume that our truncated description of $M^{(k)}$ holds. The terms
$\left<H_n(Y)|x\right>$, finally, are (for the time being) treated as $O(1)$ terms, because $Y:=\theta^{1/2}\eta$ is
a Gaussian random variable with a constant variance of $1/2$.
It remains to ask for the magnitude of $r_i(\tau)$. According to Eq.~(\ref{def_r_ab}), this is a term of
order $O(\eps^a)$ for $r_i=r_{a0}$, whereas for $r_i=r_{0b}$ or $r_i=r_{ab}$ it may be treated as term of order $O(1)$,
because in this case the value range of $r_i$ is finite and depends only on $a$ and $b$.

We now focus on a description of $M^{(1)}$, in which only terms up to order $O(\eps^3)$ are considered.
According to Eq.~(\ref{r_ab_alpha}) and the above assumptions, a function $\phi_\vecalpha$ may only give rise to terms of
order $O(\theta^j\tau^k)$ with $j+k=\ell(\vecalpha)$. Therefore, the lowest order contributions in terms of $\eps$ are of order
$O(\eps^{\ell(\vecalpha)})$. We thus may write Eq.~(\ref{series_M1}) in the form
\begin{eqnarray}
M^{(1)}(\tau,x) &=& \sum_{\ell(\vecalpha)\le 3} \!c_\vecalpha(x) \phi_\vecalpha(\tau,x)+O(\eps^4).
\end{eqnarray}

\noindent Evaluating these functions $\phi_\vecalpha$ and re-sorting terms then provides
an approximation of $M^{(1)}$ in terms of seven base functions $r_i$. Of course, the resulting coefficients of these
functions are only truncated versions of the coefficients $\lambda_i$ in Eq.~(\ref{Mk_lambda_r}). The coefficient of
$r_{1,0}\equiv\tau$, e.g., is only accurate up to order $O(\theta)$ --- but this will be sufficient for our purposes.

So far, the expectation values $\left<H_n(Y)|x\right>$ have been treated as terms of order $O(1)$. Actually, however,
this is only a lower limit for their order of magnitude. This becomes obvious by looking at the Fokker-Planck equation
of the stationary 2D-process $[X(t),\eta(t)]$, where it turns out that $\left<H_1(Y)|x\right>$ is of order $O(\eps)$
(see Appendix \ref{app_expectation_eta0}),
\begin{eqnarray}\label{mean_H1}
\left<H_1(Y)|x\right> &=& 2\theta^{1/2}\left<\eta|x\right> \;=\; -\theta^{1/2}\frac{2f(x)}{g(x)}.
\end{eqnarray}

\noindent For $n>1$, such explicit results are not available. Nevertheless, the magnitude of terms can be shown to obey
(see Appendix \ref{app_expectation_Hn})
\begin{eqnarray}\label{mean_Hn}
\left<H_n(Y)|x\right> &=& O(\theta^{n/2})\;=\;O(\eps^n).
\end{eqnarray}

\noindent With these findings, a number of terms become sufficiently small to be neglected, which leads to an approximation of
$M^{(1)}$ in terms of the function base $\{r_{0,1},r_{1,0},r_{2,0},r_{3,0}\}$.
Additionally, it turns out that terms $\left<H_n(Y)|x\right>$ with $n>1$ are no longer present in the coefficients of these
functions.

As a last step, we switch to a modified base $\{r_{0,1},r_1,r_2,r_3\}$, where the new base functions $r_i$
are linear combinations of the former ones,
\begin{eqnarray}
r_i(\tau) &:=&
\left\{\begin{array}{ll}
  r_{1,0}(\tau) - \theta\, r_{0,1}(\tau)  ,& i=1\\[.3em]
  r_{i,0}(\tau) - \theta\, r_{i-1}(\tau)  ,& i=2,3 
\end{array}\right. .
\end{eqnarray}

\noindent This base not only leads to simpler coefficients in general, but most importantly, the coefficient of
$r_{0,1}$ now becomes sufficiently small to be neglected, which leaves us with a base of only three functions.

Following the above lines, also an approximation of $M^{(2)}$ can be obtained.
As it is the case for the approximation of $M^{(1)}$, calculations are straightforward but cumbersome.
Therefore, we only give the final results here, which can be
summarized as follows. The moments $M^{(k)}$ can be approximated by
\begin{eqnarray}\label{Mk_lambda_r_final}
M^{(k)}(\tau,x) &\approx& \sum_{i=1}^3\lambda^{(k)}_i(x)\,r_i(\tau),
\end{eqnarray}
\noindent with
\begin{eqnarray}\label{def_r_i}
r_i(\tau) &=&
\left\{\begin{array}{ll}
  \tau - \theta\, (1-e^{-\tau/\theta})  ,& i=1\\[.3em]
  \frac{1}{i!}\,\tau^i - \theta\, r_{i-1}(\tau)  ,& i=2,3 
\end{array}\right. .
\end{eqnarray}

\noindent The coefficients of $r_1$ are found to be (omitting arguments)
\begin{subequations}\label{def_lambda_1_k}
\begin{eqnarray}
\lambda^{(1)}_1 &=& f +\frac{1}{2}gg' +\frac{1}{2}\theta\big\{ f'gg' -fg'g' \big\},\\
\lambda^{(2)}_1 &=& gg +\theta\big\{ f'gg-fgg' \big\}.
\end{eqnarray}
\end{subequations}

\noindent These equation will allow us to determine $f$ and $g$, once we manage to provide values for $\lambda^{(k)}_1$
and $\theta$.

\section{Parameter estimation}
\label{sec_estimation}

\noindent Now the estimation of $\lambda^{(k)}_i$ and $\theta$ can be addressed. For a given time series of $X$,
the moments $M^{(k)}$ can be estimated for a number of $N$ time-increments $\tau_\nu$ with
\begin{eqnarray}
\tau_\nu &\le& \tau_\text{max},\qquad \nu\;=\; 1,\ldots,N.
\end{eqnarray}

\noindent These estimates of $M^{(k)}$ can be fitted by means of Eq.~(\ref{Mk_lambda_r_final}) in a least-square sense.
Estimates of $\theta$ and $\lambda^{(k)}_i$ thus may be obtained by minimizing the residuals
\begin{eqnarray}
R^{(k)}(x,\veclambda^{(k)},\theta) &:=& \sum_{\nu=1}^N\Big[ M^{(k)}(x,\tau_\nu)\cr
                                   &&\qquad -\sum_{i=1}^3\lambda^{(k)}_i r_i(\tau_\nu,\theta)\Big]^2,
\end{eqnarray}

\noindent where the values $\lambda^{(k)}_i$ have been combined into the vector $\veclambda^{(k)}$ for
syntactical convenience. Additionally, the dependency of the functions $r_i$ on $\theta$ has been made explicit
by the syntax. Due to this dependency, a non-linear approach is needed for the minimization of the above residuals.

It may be noted that minimizing $R^{(k)}$ includes the estimation of $\theta$ for each value of $x$ and $k$ ---
despite the fact that $\theta$ is a constant. This is neither the most efficient nor the most accurate way for parameter
estimation.
Instead, we will estimate $\theta$ only once, based on the autocovariance $A(\tau) := \left<X(\tau)X(0)\right>$,
respectively its increments
\begin{eqnarray}
\Delta A(\tau_\nu) &:=& A(\tau_\nu)-A(0)\nonumber\\
&=& \left<\big[X(\tau_\nu)-X(0)\big]X(0)\right>.
\end{eqnarray}

\noindent Estimates of $\Delta A$ are much more accurate than estimates of $M^{(k)}$, because the latter are
based on far less data, due to the conditioning on $x$.
An approximation of $\Delta A$ that is accurate up to terms of order $O(\eps^3)$ is given by (see Appendix~\ref{app_autocorr})
\begin{eqnarray}
\Delta A(\tau) &\approx& \sum_{i=1}^3  \lambda_i r_i(\tau,\theta).
\end{eqnarray}

\noindent An estimate of $\theta$ may thus be obtained by minimizing
\begin{eqnarray}\label{residual_theta}
R(\veclambda,\theta) &:=& \sum_{\nu=1}^N\Big[ \Delta A(\tau_\nu)
                         -\sum_{i=1}^3\lambda_i r_i(\tau_\nu,\theta)\Big]^2.
\end{eqnarray}

\noindent For {\em fixed} $\theta$, the optimal values $\lambda_i^*(\theta)$
can be obtained by {\em linear regression}. This means: we can explicitly calculate
\begin{eqnarray}
\veclambda^*(\theta) &=& \operatorname*{arg\,min}_{\veclambda}\,R(\veclambda,\theta)
\end{eqnarray}

\noindent as well as the corresponding residual value $R[\veclambda^*(\theta),\theta]$. The optimal value $\theta^*$, which
corresponds to the global minimum $R[\veclambda^*(\theta^*),\theta^*]$, is thus formally given by
\begin{eqnarray}
\theta^* &=& \operatorname*{arg\,min}_{\theta}\,R[\veclambda^*(\theta),\theta].
\end{eqnarray}

\noindent In practice, $\theta^*$ may be found numerically, e.g., by using some recursive strategy to search for
the minimum of $R[\veclambda^*(\theta),\theta]$ within the interval $[0,\theta_\text{max}]$.
We safely may choose $\theta_\text{max}=\tau_\text{max}$, because,
according to our assumptions on the magnitude of $\tau$ and $\theta$, Eq.~(\ref{magnitude_theta}), we anyway
need to rely on $\theta<\tau_\text{max}$. Otherwise our series-truncation, Eq.~(\ref{Mk_lambda_r_final}),
would no longer be valid.

Once $\theta$ has been estimated, estimates of $\lambda^{(k)}_i$ become accessible by a linear regression strategy,
which allows us to explicitly calculate
\begin{eqnarray}
\veclambda^{*(k)}(x,\theta^*) &=& \operatorname*{arg\,min}_{\veclambda^{(k)}}\,R^{(k)}(x,\veclambda^{(k)},\theta^*).
\end{eqnarray}

\noindent As a last step, it remains to determine $f$ and $g$ using the estimates of $\theta$ and $\lambda^{(k)}_1$.
This is achieved by writing Eq.~(\ref{def_lambda_1_k}) in the form (omitting arguments and dropping asteriscs)
\begin{subequations}\label{fixed_point}
\begin{eqnarray}
f &=& \lambda^{(1)}_1 -\frac{1}{2}gg' -\frac{1}{2}\theta\big\{ f'gg' -fg'g' \big\},\\
g &=& \sqrt{\lambda^{(2)}_1 -\theta\big\{ f'gg-fgg' \big\}}.
\end{eqnarray}
\end{subequations}

\noindent Because $\theta$ is assumed to be small, $f$ and $g$ can be determined by a fixed-point iteration.
For given values $f^{(n)}$ and $g^{(n)}$, the right-hand sides of the above equations provide the definitions for
$f^{(n+1)}$ and $g^{(n+1)}$. However, as this requires the evaluation of spatial derivatives of $f^{(n)}$ and $g^{(n)}$,
one needs to simultaneously iterate the values at different locations $x_i$. The required derivatives can then be
estimated by some numerical differencing scheme. Appropriate starting values are provided by
$f^{(0)}(x_i)=\lambda^{(1)}_1(x_i)$ and
$g^{(0)}(x_i)=[\lambda^{(2)}_1(x_i)]^{1/2}$
. 
\section{Numerical example}
\label{sec_example}

\noindent In the following, we investigate a numerical test case, for which we use a non-symmetric, heavy-tailed,
process with multiplicative noise,
\begin{eqnarray}
\dot X &=& f(X) +g(X)\,\eta(t),\\
\dot \eta &=& -\frac{1}{\theta} \eta+\frac{1}{\theta}\xi(t),
\end{eqnarray}
\noindent with
\begin{eqnarray}
f(x) &=& -x+\frac{1}{2}x^2-\frac{1}{4}x^3,\\
g(x) &=& 1+\frac{1}{4}x^2.
\end{eqnarray}

\begin{figure}[h]
  \center{\vspace{-0.5em}\includegraphics*[width=8.5cm,angle=0]{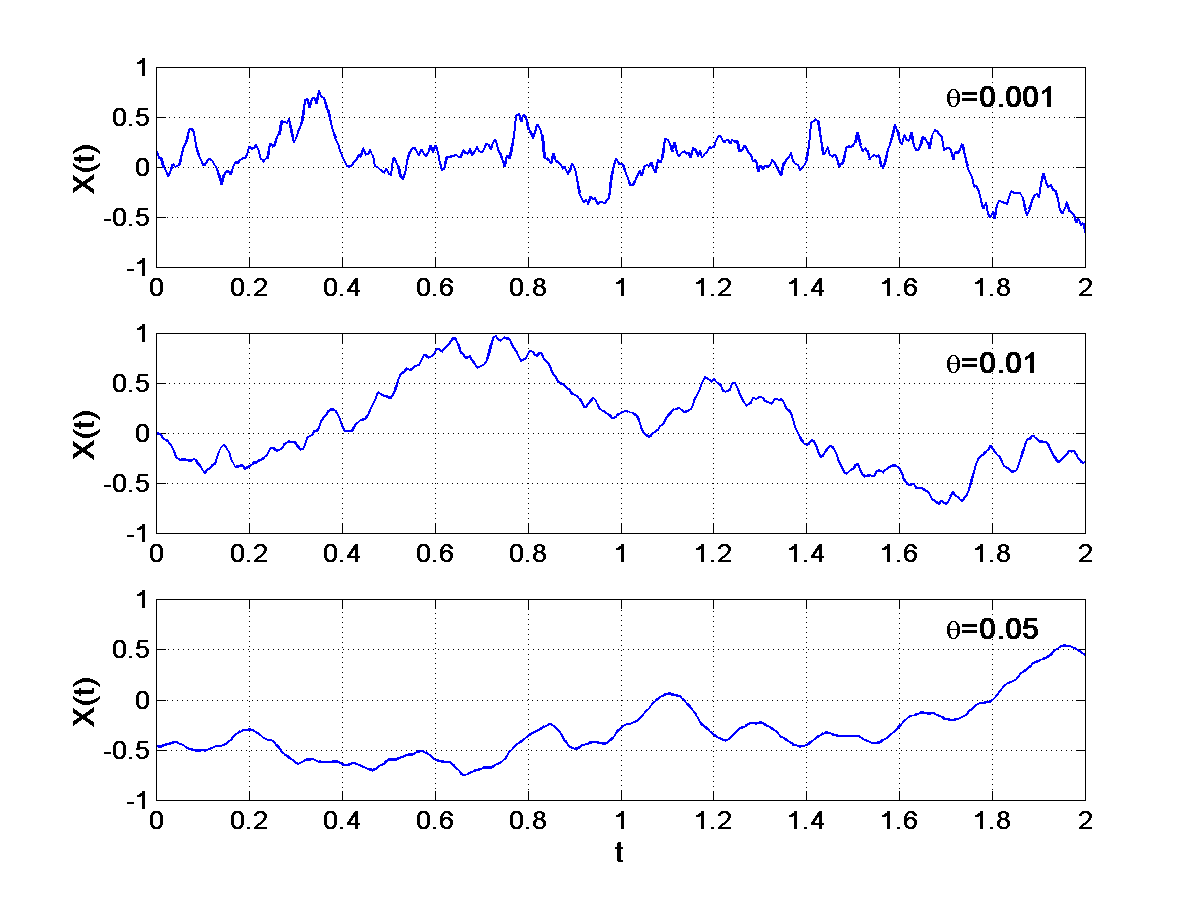}\vspace{-1.5em}}
  \caption{\protect Excerpts of the generated time series. For increasing values of $\theta$, the curves become smoother.
  }\label{fig1}
\end{figure}

\begin{figure}[h]
  \center{\vspace{-0.5em}\includegraphics*[width=8.5cm,angle=0]{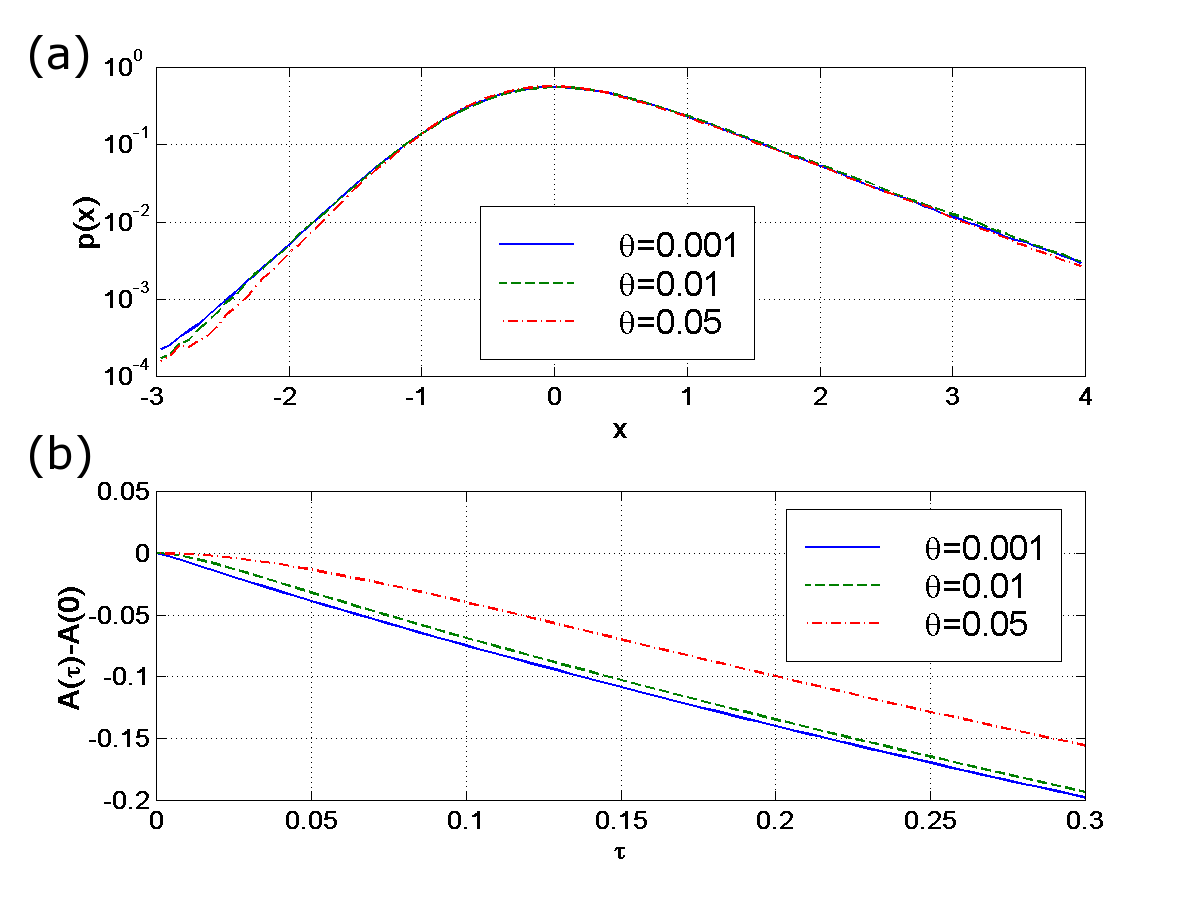}\vspace{-1.5em}}
  \caption{\protect PDFs of the experimental data  (a) and increments $\Delta A(\tau)$ of their autocorrelation (b).
  }\label{fig2}
\end{figure}

\noindent We use this system of equations to generate discrete time series of $X(t)$, consisting of $10^7$ points, using
a sampling timestep $dt\!=\!0.005$.
Integration is performed using the Euler-scheme with an internal timestep $\delta t\!=\!0.02\times\min(\theta,dt)$.
The global time-scale of $X(t)$ can be estimated from its autocorrelation function and approximately equals unity for
small values of $\theta$.
In Fig.~\ref{fig1} excerpts of the generated time series are shown for different values of $\theta$, and in
Fig.~\ref{fig2} the corresponding probability densities $p(X)$ and the increments $\Delta A$ of the autocorrelation functions
are provided. While stronger correlations of the driving noise lead to notably smoother time series, almost no effect on
the probability density can be seen.

For the analysis of a given time series, the regression functions $r_i(\tau,\theta)$ play a central role. Therefore,
we give them explicitely here again,
\begin{subequations}\label{def2_r_i}
\begin{eqnarray}
r_1(\tau,\theta) &=& \tau - \theta\, (1-e^{-\tau/\theta})\\
r_2(\tau,\theta) &=& \tau^2/2 - \theta\, r_1(\tau,\theta)\\
r_3(\tau,\theta) &=& \tau^3/6 - \theta\, r_2(\tau,\theta).
\end{eqnarray}
\end{subequations}

\noindent The actual analysis can be summarized as:
\begin{itemize}
\item[1)] Estimate the correlation time $\theta$ by non-linear fitting the increments of the autocovariance
of $X$ with the functions $r_i(\tau,\theta)$.
\item[2)] Use the estimated value $\theta^*$ to estimate the values $\lambda_1^{(k)}(x,\theta^*)$ by linear fitting
the moments $M^{(k)}(\tau,x)$ with the functions $r_i(\tau,\theta^*)$.
\item[3)] Use the estimated values $\lambda_1^{*(k)}(x,\theta^*)$ to calculate estimates for $f$ and $g$ using
Eq.~(\ref{fixed_point}).
\end{itemize}

\noindent These steps will now be detailed. We first consider the estimation of the correlation time $\theta$.
As mentioned in Sec.~\ref{sec_estimation}, an estimate $\theta^*$
may be found by minimizing the residual $R[\veclambda^*(\theta),\theta]$, where the vector $\veclambda^*(\theta)$ is
obtained from a linear fit of $\Delta A(\tau)$ using the functions $r_i(\tau,\theta)$.
To find the minimum of $R$ in an interval $[\theta_\text{min},\theta_\text{max}]$, we use a recursive strategy.
First, the residual is evaluated for a number of equidistant values $\theta_i$ covering the whole interval.
Next, the interval is narrowed and repositioned such that it only covers the vicinity of the value $\theta_i^*$, for
which the residual was found to be smallest. These steps can now be repeated until the desired numerical accuracy
is reached.

In our example, we first use the values $\Delta A(\nu\, dt)$ with $1\le\nu\le 60$ for the fits. This corresponds to a maximum
time increment $\tau_\text{max}=0.3$. Since we use a truncated series representation for the description of $\Delta A$,
the value of $\tau_\text{max}$ affects the {\em systematic} errors of the fits and should be choosen as small as possible.
Therefore, once we have calculated $\theta^*$ with $\tau_\text{max}=0.3$, we restrict the maximum increment to
$\tau^*_\text{max}=\sqrt{\theta^*}$, which is consistent with our assumptions on the magnitude of terms, and repeat the
calculation of $\theta^*$.

\begin{figure}[h]
  \center{\vspace{-0.4em}\includegraphics*[width=8.5cm,angle=0]{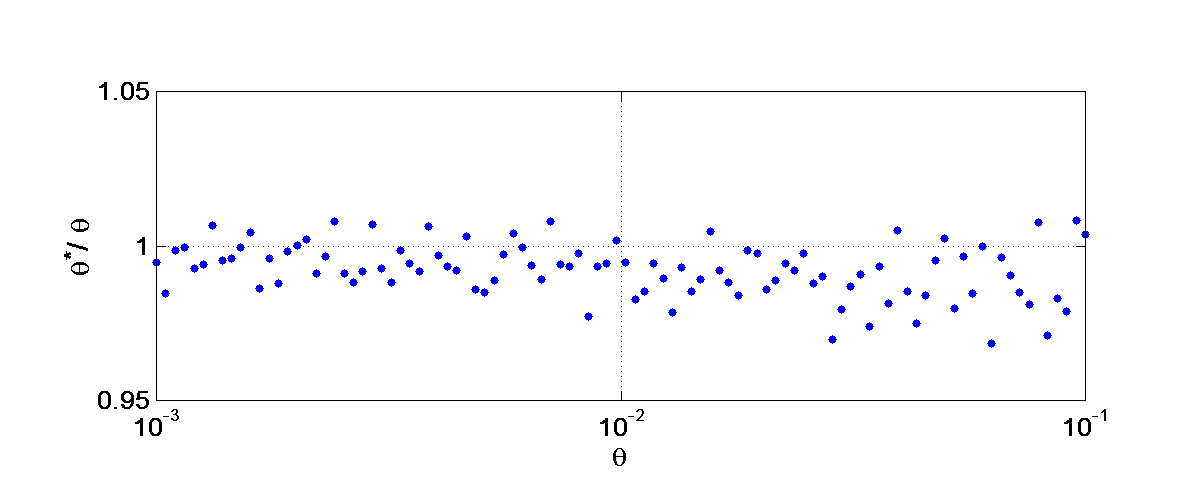}\vspace{-1.5em}}
  \caption{\protect Ratio $\theta^*/\theta$ of estimated and true correlation time.
  }\label{fig3}
\end{figure}

Estimates for $\theta$ that are obtained by following this strategy are shown in Fig.~\ref{fig3} for the range
$0.001\le\theta\le 0.1$. Even if $\theta$ seems to be slightly underestimated for $\theta>0.01$, the overall accuracy
is quite good.

\begin{figure}[h]
  \center{\vspace{-0.4em}\includegraphics*[width=8.5cm,angle=0]{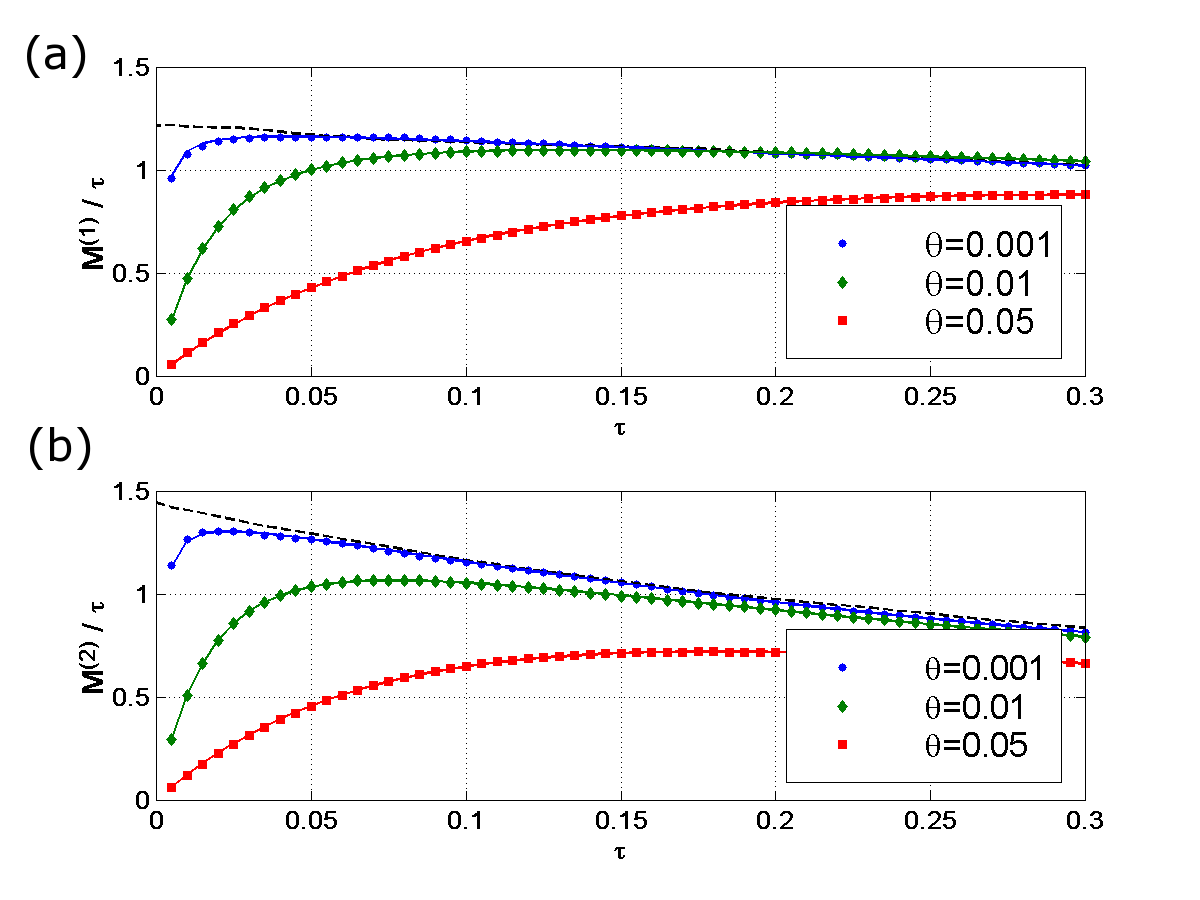}\vspace{-1.5em}}
  \caption{\protect Estimates of $M^{(1)}$ (a) and $M^{(2)}$ (b), obtained for a bin centered at $x=-0.9$. Estimated 
           values are shown as symbols and the corresponding fits as solid lines. Additionally, the moments in the limit
           $\theta\to 0$ are indicated by dashed lines.
  }\label{fig4}
\end{figure}

With an estimate $\theta^*$ at hand, the coefficients $\lambda_1^{(k)}$ are obtained from linear fits
of the moments $M^{(k)}$ using the functions $r_i(\tau,\theta^*)$. For the estimation of these moments, we use a binning
approach, where the range $-2\le x\le 3$ is divided into 25 bins. For each bin we estimate the values $M^{(k)}(\nu\, dt,x)$
with $1\le\nu\le 60$ from the data. Here, $x$ is taken to be the position of the bin-center. In Fig.~\ref{fig4}, estimated
values and resulting fits of $M^{(k)}$ at $x=-0.9$ are shown for different values of $\theta$. The values obtained from
the data can excellently be fitted with the functions $r_i$. The mean error is only about $2.5\times 10^{-4}$.

\begin{figure}[H]
  \center{\vspace{-0.4em}\includegraphics*[width=8.5cm,angle=0]{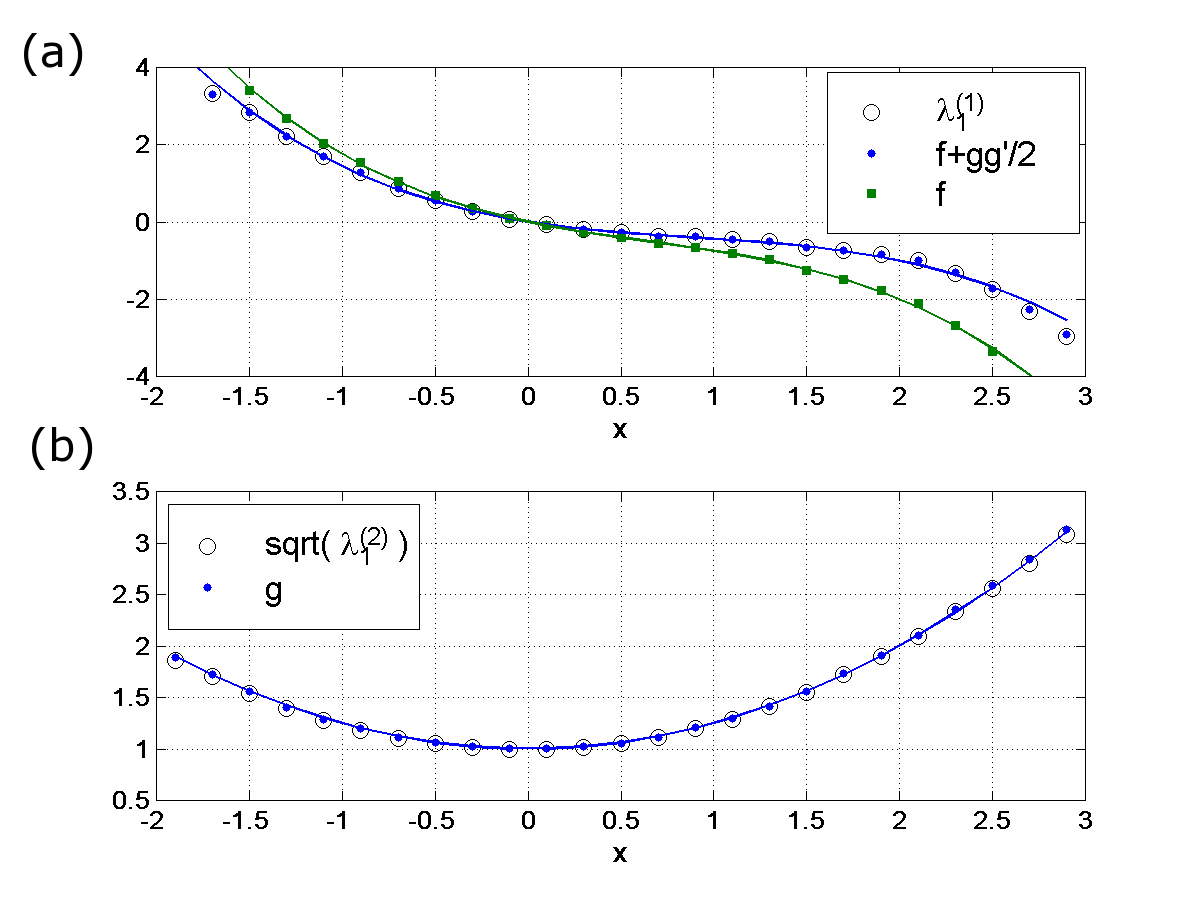}\vspace{-1.5em}}
  \caption{\protect Estimates of $\lambda_1^{(1)}$, $f$ and $f+gg'/2$ (a) and $[\lambda_1^{(2)}]^{1/2}$ and $g$ (b)
           for $\theta=0.01$. Estimated values are shown as symbols and the true functions as solid lines.
  }\label{fig5}
\end{figure}

Finally, we consider the estimates of the coefficients $\lambda_1^{(k)}$ and of the functions $f$, $f+gg'/2$ and
$g$. We do not show the results for $\theta=0.001$, because these would look almost identical to the results for
$\theta=0.01$, which are presented in Fig.~\ref{fig5}. Here, we find that the estimates of $f$ and $g$ are in very good
accordance with the true values. Additionally, it shows that --- for the given value of $\theta$ --- the values of
$\lambda_1^{(1)}$ and $[\lambda_1^{(2)}]^{1/2}$ are almost identical to the values of $f+gg'/2$ and g, respectively.
But this will change, when larger values of $\theta$ are considered.

\begin{figure}[h]
  \center{\vspace{-0.5em}\includegraphics*[width=8.5cm,angle=0]{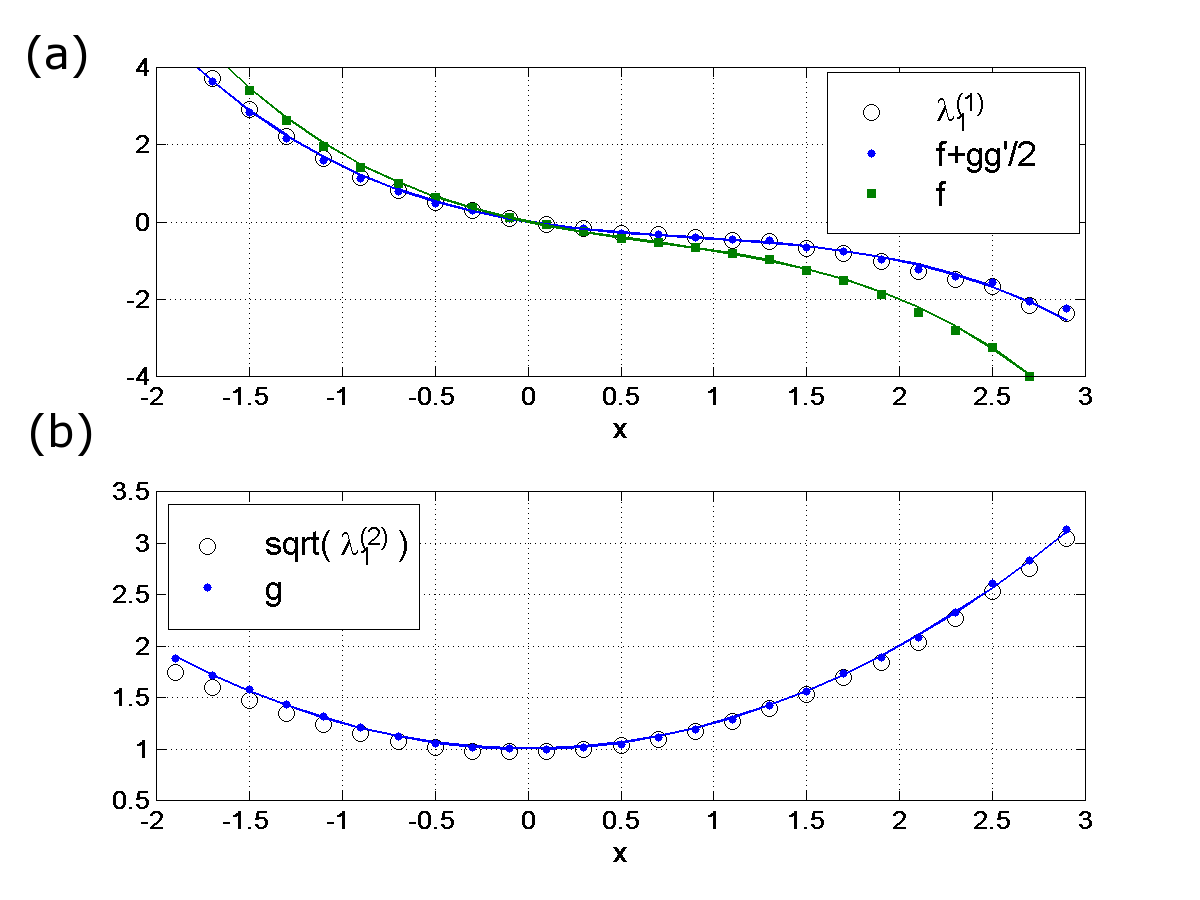}\vspace{-1.5em}}
  \caption{\protect Estimates of $\lambda_1^{(1)}$, $f$ and $f+gg'/2$ (a) and $[\lambda_1^{(2)}]^{1/2}$ and $g$ (b)
           for $\theta=0.05$. Estimated values are shown as symbols and the true functions as solid lines.  }\label{fig6}
\end{figure}

In Fig.~\ref{fig6} the results for $\theta=0.05$ are shown, where $[\lambda_1^{(2)}]^{1/2}$ clearly deviates from $g$.
However, since we account for this deviation by Eq.~(\ref{fixed_point}), we still obtain accurate estimates for $f$ and $g$.

\begin{figure}[h]
  \center{\vspace{-0.5em}\includegraphics*[width=8.5cm,angle=0]{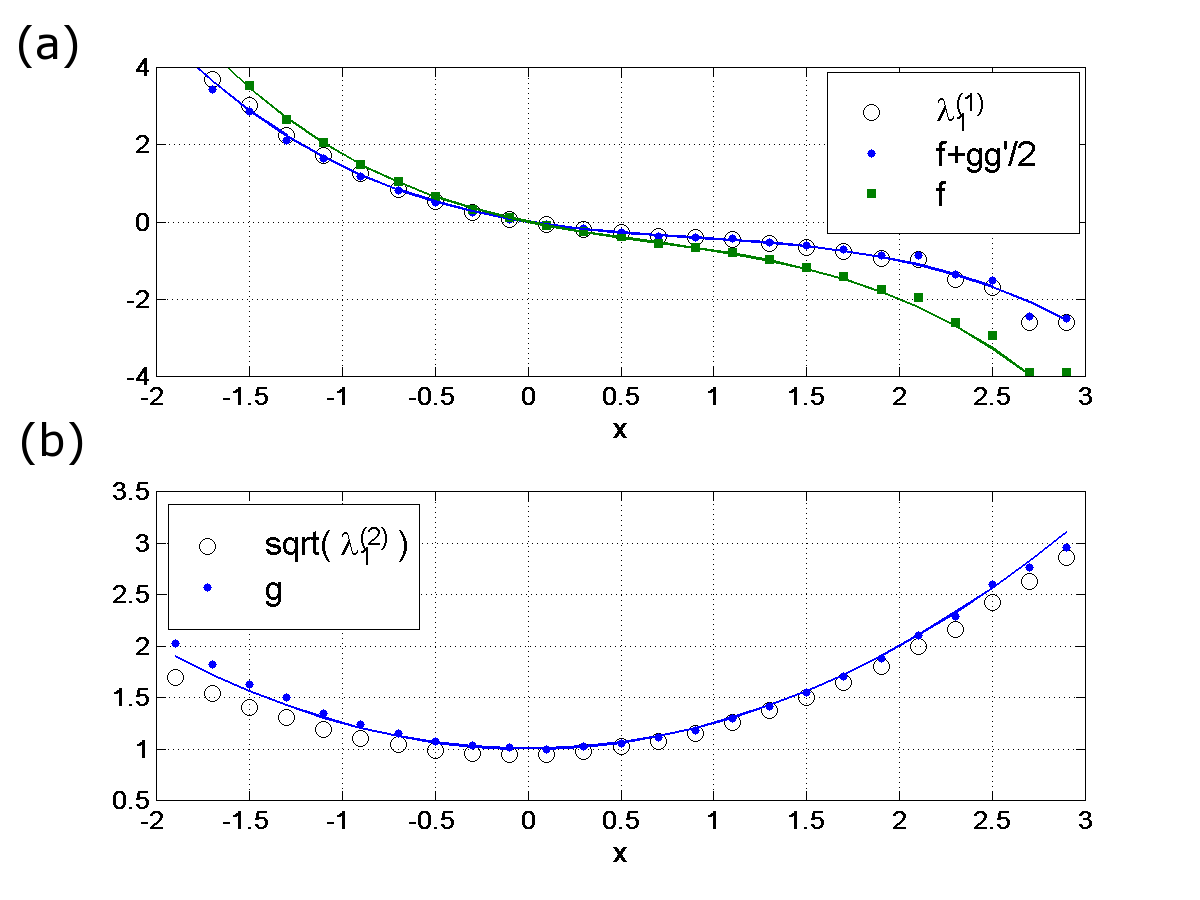}\vspace{-1.5em}}
  \caption{\protect Estimates of $\lambda_1^{(1)}$, $f$ and $f+gg'/2$ (a) and $[\lambda_1^{(2)}]^{1/2}$ and $g$ (b)
           for $\theta=0.1$. Estimated values are shown as symbols and the true functions as solid lines.
  }\label{fig7}
\end{figure}

To also show the limitations of our approach, the results for $\theta=0.1$ are presented in Fig.~\ref{fig7}.
Here, the estimates become less accurate outside the range $-1<x<2$. Especially for $x>2$ the estimates of $\lambda_1^{(1)}$
now show fluctuations that hamper the estimation of the spatial derivatives that are needed to calculate $f$ and $g$
using Eq.~(\ref{fixed_point}).

\section{Conclusions}

\noindent A parameter-free approch has been developed that allows for the analysis of a stochastic process $X(t)$ that
is driven by exponentially correlated, Gaussian noise. This analysis is purely based on the moments of the conditional
increments of $X$ and provides estimates for the drift- and diffusion-functions of the process as well as for the correlation
time $\theta$ of the driving noise.

It should be noted that we use a perturbative approach, where $\theta$ is assumed to be small as compared to the
characteristic time-scale of $X$. Actually, the method presented in this paper is accurate up to first order terms in
$\theta$. In principle, however, also higher order approximations are possible.
The method may be seen as generalization of the direct estimation method \cite{friedrich11}, which also formally is
recovered in the limit $\theta\to 0$.

The applicability and accuracy of our approach has been demonstrated by a numerical example, where reasonable accurate
results are obtained even for values of $\theta$ as large as ten percent of the global time-scale.
For smaller values of $\theta$, the results are (aside finite-size fluctuations) close to exact.

The presented approach is straightforward to implement and neither demanding with regard to memory nor to CPU power. An
analysis of a series of $10^7$ values is performed within a few seconds on a standard desktop PC.

It would be interesting to also apply our approach to an analysis of turbulent velocity increments as described in
\cite{friedrich97,renner01}, where the description of the increments is approximated by a Markov-process in scale. The observed
conditional moments of the process-increments show a striking similarity to the moments $M^{(k)}$ [Fig.~\ref{fig4}] of a
process that is driven by Ornstein-Uhlenbeck noise. However, since the drift- and diffusion functions for this problem
clearly depend on scale, our approach first needs to be extended to non-stationary processes, which is a task
for the future.


\appendix
\section{Products of integrals $J_\vecalpha$}
\label{app_products_j_integrals}

\noindent According to Eq.~(\ref{def_J_alpha}), the integrals $J_\vecalpha$ are defined as
\begin{eqnarray}
J_{(\alpha_1,\ldots,\alpha_n)}(\tau) &:=&
\int_{s_n=0}^{\tau}
\int_{s_{n\!-\!1}=0}^{s_n}\!\cdots
\int_{s_1=0}^{s_2} \nonumber\\[.3em]
&&\times
\,dZ_{\alpha_1}(s_1)\cdots dZ_{\alpha_n}(s_n) ,
\end{eqnarray}
\noindent with
\begin{eqnarray}
dZ_j(s) &:=& \left\{\begin{array}{ll}
ds & \quad, j=0 \\
dV(s) & \quad, j=1
\end{array}\right. ,
\end{eqnarray}

\noindent There are a number of obvious solutions, like
\begin{eqnarray}
J_{(0)}(\tau) &=& \tau,\qquad J_{(1)}(\tau) \;=\; V(\tau).
\end{eqnarray}

\noindent Additionally, it is possible to express a product of two integrals by a sum of single integrals: According to the
above definition, an integral $J_\vecalpha$ may be written as
\begin{eqnarray}
J_{\vecalpha}(\tau) &:=&
\int_{s=0}^{\tau} J_{\vecalpha^{\!-}}(s)\, dZ_{\alpha_n}(s),
\end{eqnarray}

\noindent where the syntax $(\alpha_1,\ldots,\alpha_n)^-\!:=\!(\alpha_1,\ldots,\alpha_{n-1})$ has been introduced.
The differential increment of $J_\vecalpha$ is thus given by
\begin{eqnarray}
dJ_\vecalpha(\tau) &=& J_{\vecalpha^{\!-}}(\tau)\, dZ_{\alpha_n}(\tau).
\end{eqnarray}

\noindent For the differential increment of a product $J_\vecalpha J_\vecbeta$ one finds
\begin{eqnarray}
d(J_\vecalpha J_\vecbeta) &=& J_\vecalpha\,dJ_\vecbeta+J_\vecbeta\,dJ_\vecalpha.
\end{eqnarray}

\noindent In integrated form, we obtain (assuming $\vecbeta$ to have $m$ components)
\begin{eqnarray}
J_\vecalpha(\tau) J_\vecbeta(\tau) &=&
  \int_{s=0}^{\tau} J_{\vecalpha}(s) J_{\vecbeta^{\!-}}(s)\, dZ_{\beta_m}(s) \cr
&& +\int_{s=0}^{\tau} J_{\vecalpha^{\!-}}(s) J_{\vecbeta}(s)\, dZ_{\alpha_n}(s).
\end{eqnarray}

\noindent This equation can now be applied recursively to the products within the integrals. In the end, this leads to
a sum of integrals $J_\vecgamma$ with index-vectors of length $n+m$,
\begin{eqnarray}
J_\vecalpha J_\vecbeta &=& \sum_{\vecgamma\in{\cal M}(\vecalpha,\vecbeta)} J_\vecgamma.
\end{eqnarray}

\noindent Actually, the vectors $\vecgamma\in{\cal M}(\vecalpha,\vecbeta)$ represent all ${n+m\choose n}$ possibilities
to mix the indices of $\vecalpha$ and $\vecbeta$ while keeping their relative ordering, This means, the position of
$\alpha_i$ in $\vecgamma$ must always preceed that of $\alpha_{i+1}$, and the same must hold for the components of
$\vecbeta$. As an example, one obtains
\begin{eqnarray}
{\cal M}\left[(\alpha_1,\alpha_2),(\beta_1)\right] &=&
\big\{ (\beta_1,\alpha_1,\alpha_2), (\alpha_1,\beta_1,\alpha_2),\cr&& \;\;(\alpha_1,\alpha_2,\beta_1) \big\}.
\end{eqnarray}

\noindent With this rule at hand, one finds, e.g.,
\begin{subequations}
\begin{eqnarray}
J_{(0)}J_{(0)} &=& 2J_{(0,0)},\\
J_{(0)}J_{(0,0)} &=& 3J_{(0,0,0)},\\
&\vdots&\cr
J_{(1)}J_{(1)} &=& 2J_{(1,1)},\\
J_{(1)}J_{(1,1)} &=& 3J_{(1,1,1)},\\
&\vdots&\nonumber
\end{eqnarray}
\end{subequations}
\noindent which implies (assuming a vector of length $n$)
\begin{eqnarray}
J_{(0,\ldots,0)}(\tau) &=& \frac{1}{n!}[J_{(0)}(\tau)]^n \;=\;\frac{1}{n!}\tau^n,\\
J_{(1,\ldots,1)}(\tau) &=& \frac{1}{n!}[J_{(1)}(\tau)]^n \;=\;\frac{1}{n!}[V(\tau)]^n.
\end{eqnarray}

\noindent One may also derive relations like
\begin{eqnarray}
J_{(0)}J_{(1)} &=& J_{(1,0)}+J_{(0,1)},\label{sum_otto_integrals1}\\
J_{(0)}^2J_{(1)} &=& 2[J_{(1,0,0)}+J_{(0,1,0)}+J_{(0,0,1)}],\label{sum_otto_integrals2}\\
J_{(0)}J_{(1)}^2 &=& 2[J_{(0,1,1)}+J_{(1,0,1)}+J_{(1,1,0)}],\label{sum_otto_integrals3}\\
J_{(1,0)}J_{(1,0)} &=& 2J_{(1,0,1,0)}+4J_{(1,1,0,0)}.
\end{eqnarray}

\section{Correlation functions of $u(t)$}
\label{app_correlations_u}

\noindent Using the definition of $u(t)$ [Eq.~(\ref{def_u})], the definition of $C$ [Eq.~(\ref{def_C})] reads
\begin{eqnarray}\label{C_in_terms_of_xi}
C(t_1,\ldots,t_n) &=& \Big< \frac{1}{\theta^n}\int_{s_1=0}^{t_1}\cdots\int_{s_n=0}^{t_n}\cr&&\times
                     \prod_{i=1}^{n}\e{(s_i-t_i)}\xi(s_i)\cr&&\times
                     \,ds_1\cdots ds_n\Big>\cr
                 &=& \frac{1}{\theta^n}\int_{s_1=0}^{t_1}\cdots\int_{s_n=0}^{t_n}
                     \big<\xi(s_1)\cdots\xi(s_n)\big>\cr&&\times \prod_{i=1}^{n}\e{(s_i-t_i)}
                     \,ds_1\cdots ds_n.
\end{eqnarray}

\noindent As $\xi(t)$ is Gaussian white noise, the expectation values $\left<\xi(t_1)\cdots\xi(t_n)\right>$ are
well known. For odd values of $n$ they are vanishing,
\begin{eqnarray}\label{C_odd_vanishes}
\left<\xi(t_1)\cdots\xi(t_n)\right> &=& 0,\qquad  n=2k+1.
\end{eqnarray}

\noindent All even correlations can be expressed in terms of two-point correlations. For $n=2k$, this
leads to a sum of $k$-fold productes of delta functions. This sum contains $1\times3\times\cdots\times(n\!-\!1)$
terms, which is the number of possibilities to permutate the function arguments of such a product when only
distinguishable functions are allowed (functions may be indistinguishable due to the symmetry of the delta function or
due to the commutativity of multiplication). Up to $n=4$ this reads
\begin{eqnarray}
\left<\xi(t_1)\xi(t_2)\right> &=& \delta(t_1\!-\!t_2),\\
\left<\xi(t_1)\xi(t_2)\xi(t_3)\xi(t_4)\right> &=& \delta(t_1\!-\!t_2)\delta(t_3\!-\!t_4)\cr&&
                                                 +\delta(t_1\!-\!t_3)\delta(t_2\!-\!t_4)\cr&&
                                                 +\delta(t_1\!-\!t_4)\delta(t_2\!-\!t_3).
\end{eqnarray}

\noindent According to Eq.~(\ref{C_in_terms_of_xi}), it follows immediately that $C$ also vanishes for odd values of $n$,
\begin{eqnarray}
C(t_1,\ldots,t_n) &=& 0,\qquad  n=2k+1.
\end{eqnarray}

\noindent For $n=2$ one finds
\begin{eqnarray}
\label{def_Ct1t2}
C(t_1,t_2) &=& \frac{1}{\theta^2}\int_{s_1=0}^{t_1}\int_{s_2=0}^{t_2}\delta(s_1\!-\!s_2) \nonumber\\[.3em]
           &&\times \e{(s_1+s_2-t_1-t_2)}\,ds_1\,ds_2 \nonumber\\[.3em]
           &=& \frac{1}{\theta^2}\int_{s=0}^{\min(t_1,t_2)}\e{(2s-t_1-t_2)}\,ds \nonumber\\[.3em]
           &=&
\left\{\begin{array}{ll}
  \frac{\e{(t_1-t_2)} - \e{(-t_1-t_2)}}{2\theta}  ,& t_1 \le   t_2\\[.3em]
  \frac{\e{(t_2-t_1)} - \e{(-t_2-t_1)}}{2\theta}  ,& t_1 > t_2 
\end{array}\right. .\label{twopoint_eta}
\end{eqnarray}

\noindent The higher order correlation functions of $u(t)$ can be expressed in terms of two-point correlations like
in the case of Gaussian white noise. This can be seen when inserting the expressions for
$\left<\xi(t_1)\cdots\xi(t_{2k})\right>$ into Eq.~(\ref{C_in_terms_of_xi}). For each of the $k$-fold products of the delta
function, the integral factorizes into a $k$-fold product of integrals of the form of Eq.~(\ref{def_Ct1t2}). Therefore,
the structure of the correlation functions of $\xi(t)$ directly translates to that of $u(t)$. One finds
\begin{eqnarray}
C(t_1,t_2,t_3,t_4) &=& C(t_1,t_2)C(t_3,t_4)\cr
                   &&+C(t_1,t_3)C(t_2,t_4)\cr
                   &&+C(t_1,t_4)C(t_2,t_3),\\
&\vdots&\nonumber
\end{eqnarray}

\section{Explicit example for $\phi_\vecalpha$}
\label{app_example_phialpha}

\noindent To provide an explicit example for the calculation of $\phi_\vecalpha$, the case $\vecalpha=(1,0,1)$ is considered.
Equation~(\ref{def_phi_alpha_explicit}) then reads
\begin{eqnarray}
\phi_{(1,0,1)}(\tau,x) = \int_{\Omega(\tau)}\!\!\! C_\eta(s_1,s_3,x) \,ds_1 ds_2 ds_3,
\end{eqnarray}

\noindent where we again denote the integration domain $0\le s_1 \le s_2 \le s_3 \le \tau$ by $\Omega(\tau)$ and $C_\eta$ is
defined according to Eq.~(\ref{def_C_eta}).

By expressing $\eta$ in terms of $Y$ and $u$ [Eqs. (\ref{split_eta}), (\ref{def_Y}) and (\ref{def_u})], the two-point
correlation $C_\eta(s_1,s_3,x)$ can be expressed in terms of moments of $Y$ and correlation functions uf $u$, which are
denoted by $C$ and defined according to Eq.~({\ref{def_C})
\begin{eqnarray}
C_\eta(s_1,s_3,x) \!&=&
\big<Y^2|x\big>\frac{e^{-(s_1+s_3)/\theta}}{\theta}\nonumber\\
   \!&&+\big<Y|x\big>\frac{e^{-s_1/\theta}}{\sqrt{\theta}}C(s_3)\nonumber\\
   \!&&+\big<Y|x\big>\frac{e^{-s_3/\theta}}{\sqrt{\theta}}C(s_1)\nonumber\\
   \!&&+C(s_1,s_3).
\end{eqnarray}

\noindent Taking into account that, according to Eq.~(\ref{C_odd_vanishes}), the functions $C(s_1)$ and $C(s_3)$ are vanishing,
this leaves us with (omitting index vector and arguments for $\phi$)
\begin{eqnarray}
\phi
  &=& \big<Y^2|x\big> \int_{\Omega(\tau)}\!\!\! \frac{e^{-(s_1+s_3)/\theta}}{\theta} \,ds_1 ds_2 ds_3 \nonumber\\
  &&  +\int_{\Omega(\tau)}\!\!\! C(s_1,s_3) \,ds_1 ds_2 ds_3.
\end{eqnarray}

\noindent Inserting Eq.~(\ref{def_Ct1t2}) and taking into account that $s_1\le s_3$ holds within the integration domain, this
becomes
\begin{eqnarray}
\phi
  &=& \big<Y^2|x\big> \int_{\Omega(\tau)}\!\!\! \frac{e^{-(s_1+s_3)/\theta}}{\theta} \,ds_1 ds_2 ds_3 \nonumber\\
  &&  +\int_{\Omega(\tau)}\!\!\! \frac{\e{(s_1-s_3)}}{2\theta} \,ds_1 ds_2 ds_3 \nonumber\\
  &&  -\int_{\Omega(\tau)}\!\!\! \frac{\e{(-s_1-s_3)}}{2\theta} \,ds_1 ds_2 ds_3,
\end{eqnarray}

\noindent which evaluates to
\begin{eqnarray}
\phi
  &=& \left<Y^2|x\right> a_0(\tau) + a_1(\tau)
\end{eqnarray}
\noindent with
\begin{subequations}
\begin{eqnarray}
a_0(\tau) &=& \frac{1}{2}\theta^2 (1-\e{-2\tau}) -\tau\theta \e{-\tau},\\
a_1(\tau) &=& \frac{1}{2}\tau\theta  -\theta^2 (1-\e{-\tau})\nonumber\\
          && -\frac{1}{4}\theta^2 (1-\e{-2\tau}) +t\theta \e{-\tau}.
\end{eqnarray}
\end{subequations}

\noindent In terms of the Hermite Polynomials $H_0(Y):=1$ and $H_2(Y):=4Y^2-2$ this can be re-written as
\begin{eqnarray}
\phi &=& \left<H_2(Y)|x\right> b_0(\tau) + \left<H_0(Y)|x\right> b_1(\tau).
\end{eqnarray}
\noindent with
\begin{subequations}
\begin{eqnarray}
b_0(\tau) &=& \frac{1}{8}\theta^2 (1-\e{-2\tau}) -\frac{1}{4}\tau\theta \e{-\tau},\\
b_1(\tau) &=& \frac{1}{2}\tau\theta  -\theta^2 (1-\e{-\tau})+\frac{1}{2}t\theta \e{-\tau}.
\end{eqnarray}
\end{subequations}

\section{Expectation value of $\eta\big|_x$}
\label{app_expectation_eta0}

\noindent Equations~(\ref{evolution_X}) and (\ref{evolution_eta}) describe a Markov process in two
dimensions. Using $x$ and $s$ to denote the phase-space variables of $X$ and $\eta$, the Kramers--Moyal coefficients of
the corresponding Fokker--Planck equation are given by
\begin{eqnarray}
\vecDDrift(x,s) &=& \begin{bmatrix} f(x)+g(x)s \\ -s/\theta\end{bmatrix},\\
\vecDDiff(x,s)  &=& \begin{bmatrix} 0 & 0 \\ 0 & 1/\theta^2\end{bmatrix}.
\end{eqnarray}

\noindent The Fokker--Planck equation thus reads
\begin{eqnarray}
\p_t p(x;s) &=& -\p_x\big\{p(x;s)[f(x)+g(x)s]\big\}\nonumber\\[.2em]
               &&+\p_s\big\{p(x;s)s/\theta\big\}\cr
               &&+\frac{1}{2}\p_s^2\big\{p(x;s)/\theta^2\big\}.
\end{eqnarray}

\noindent Integrating with respect to $s$ then gives [using $p(x;s)\!=\!p(x)p(s|x)$
and $\int_s\!s\, p(s|x)\!=\!\left<\eta|x\right>$]
\begin{eqnarray}
\p_t p(x) &=& -\p_x\big\{ p(x)[ f(x)+g(x)\left<\eta|x\right>] \big\}\nonumber\\[.2em]
          &=& -\p_x j(x),
\end{eqnarray}

\noindent where $j$ denotes the probability flux. For the stationary process we have $\p_t p(x)=0$, implying
a constant flux. For natural boundary conditions (vanishing flux and density at $|x|\to\infty$) this implies
$j\equiv 0$ and thus
\begin{eqnarray}
\left<\eta|x\right> &=& -\frac{f(x)}{g(x)}.
\end{eqnarray}

\section{Expectation values $\big<H_n(Y)|x\big>$}
\label{app_expectation_Hn}

\noindent In terms of $Y(t)\equiv\sqrt{\theta}\eta(t)$ our evolution equations [Eqs.~(\ref{evolution_X}) and
(\ref{evolution_eta})] read
\begin{eqnarray}
\fracpp{t} X &=& f(X) +\frac{1}{\sqrt{\theta}}g(X)\,Y,\\
\fracpp{t} Y &=& -\frac{1}{\theta} Y+\frac{1}{\sqrt{\theta}}\,\xi(t).
\end{eqnarray}

\noindent The corresponding Fokker-Planck equation for the stationary process may then be written as
\begin{eqnarray}\label{def_FokkerPlanck_XY}
0 &=& -\fracpp{x}\Big\{p(x;y)\big[\theta f(x)+\sqrt{\theta}\, g(x)\,y\big]\Big\}\cr
  &&  +\fracpp{y}\big[p(x;y)\,y\big] +\frac{1}{2}\frac{\p^2}{\p y^2}\,p(x;y),
\end{eqnarray}

\noindent where $p(x;y)$ denotes the stationary joint PDF of $X(t)$ and $Y(t)$. Actually, this density also depends
on the parameter $\theta$, but we will not make this explicit by the syntax. In the following, we express
$p(x;y)$ by a Hermite expansion of the form
\begin{eqnarray}\label{def_expansion}
p(x;y) &=& p_0(x)\sum_{i=0}^\infty \theta^{i/2} c_i(x,\theta)\, H_i(y)\, G(y),
\end{eqnarray}

\noindent with
\begin{eqnarray}
\label{def_c_i}
c_i(x,\theta) &:=& \sum_{j=0}^\infty \theta^j c_{i,j}(x),\\
G(y) &:=& \frac{1}{\sqrt{\pi}}\,e^{-y^2},\\
H_i(y) &:=& (-1)^i\frac{1}{G(y)}\, \frac{\p^i}{\p y^i} G(y).
\end{eqnarray}

\noindent The function $p_0(x)$, finally, denotes the density of $X(t)$ in the limit $\theta\to 0$. In this limit,
Eq.~(\ref{evolution_X}) becomes $\dot X=f+g\xi$ (to be interpreted in the Stratonovich sense). The stationary density
of $X$ can then be calculated from the corresponding Fokker-Planck equation
$0=\p_x[fp_0-\frac{1}{2}g\p_x(gp_0)]$, leading to
\begin{eqnarray}\label{def_p0}
p_0(x) &:=& \lim_{\theta\to 0} p(x)\nonumber\\
&=& \frac{N}{g(x)}\exp\left(\int_{-\infty\!}^x \frac{2f(s)}{g^2(s)}\,ds \right),
\end{eqnarray}

\noindent where $N$ is a normalization constant.

As the Hermite polynomials $H_n(y)$ are orthogonal under the weight $G(y)$, i.e.,
\begin{eqnarray}
\int_{-\infty\!}^\infty H_n(y)\,H_m(y)\,G(y)\,dy &=& 2^n n!\, \delta_{nm},
\end{eqnarray}

\noindent one first finds from Eq.~(\ref{def_expansion})
\begin{eqnarray}
\int_{-\infty\!}^\infty H_n(y)\,p(x;y)\,dy &=& \theta^{n/2}2^n n! \cr&&\times p_0(x) c_n(x,\theta) .
\end{eqnarray}

\noindent Using $p(x;y)=p(x)p(y|x)$ and $\int_yF(y)p(y|x)=\left<F(Y)|x\right>$, the left-hand side of this equation may
be rewritten to obtain
\begin{eqnarray}
p(x)\left<H_n(Y)|x\right> &=& \theta^{n/2}2^n n!\,p_0(x) c_n(x,\theta) .
\end{eqnarray}

\noindent For the case $n=0$, one finds (because of $H_0(y)\equiv 1$)
\begin{eqnarray}
p(x) &=& p_0(x)c_0(x,\theta).
\end{eqnarray}

\noindent Together with Eq.~(\ref{def_c_i})
and $p_0(x)=\lim_{\theta\to 0}p(x)$, this provides us with the value of $c_{0,0}$,
\begin{eqnarray}
c_{0,0}(x) &=& \lim_{\theta\to 0} c_0(x,\theta) \;=\; \lim_{\theta\to 0}\frac{p(x)}{p_0(x)} \;=\; 1.
\end{eqnarray}

\noindent For $n=1$, a relation between the coefficients $c_{0,i}$ and $c_{1,i}$ can be obtained by using
$\left<H_1(Y)|x\right>=-2\sqrt{\theta}f/g$ [Eq.~(\ref{mean_H1})] and $p=p_0c_0$. One first finds
\begin{eqnarray}\label{relation_c1_c0}
c_1(x,\theta) &=& -\frac{f(x)}{g(x)}c_0(x,\theta),
\end{eqnarray}

\noindent and as this equation holds for arbitrary $\theta$, Eq.~(\ref{def_c_i}) implies
\begin{eqnarray}\label{relation_c1i_c0i}
c_{1,i}(x) &=& -\frac{f(x)}{g(x)}c_{0,i}(x).
\end{eqnarray}

\noindent In the general case one obtains
\begin{eqnarray}
\left<H_n(Y)|x\right> &=& \theta^{n/2}2^n n!\,\frac{c_n(x,\theta)}{c_0(x,\theta)}.
\end{eqnarray}

\noindent Assuming the coefficients $c_{i,j}$ to be of order $O(1)$ then yields
\begin{eqnarray}
\frac{c_n(x,\theta)}{c_0(x,\theta)} = \frac{c_{n,0}(x)+O(\theta)}{1+O(\theta)} = c_{n,0}(x)+O(\theta),
\end{eqnarray}

\noindent which implies $\left<H_n(Y)|x\right>=O(\theta^{n/2})$, as claimed by Eq.~(\ref{mean_Hn}).

It remains to be shown,
however, that there exists a set of finite coefficients $c_{i,j}$, for which
Eq.~(\ref{def_expansion}) is a solution of the Fokker-Planck equation as specified by Eq.~(\ref{def_FokkerPlanck_XY}).
In order to calculate these coefficients, we insert Eq.~(\ref{def_expansion}) into Eq.~(\ref{def_FokkerPlanck_XY}).
Using the well known relations
\begin{eqnarray}
y\, H_n(y) &=& n\,H_{n-1}(y)+\frac{1}{2}H_{n+1}(y)
\end{eqnarray}
\noindent and
\begin{eqnarray}
\frac{\p}{\p y}[H_n(y)\,G(y)] &=& -H_{n+1}(y)\,G(y),
\end{eqnarray}

\noindent this first leads to (omitting arguments)
\begin{eqnarray}
0 &=& -\fracpp{x}\Big\{
f p_0\theta\sum_{i=0}^\infty\theta^{i/2}c_iH_iG \cr&&
+g p_0\theta^{1/2}\sum_{i=0}^\infty\theta^{i/2}c_i[ iH_{i-1}+\frac{1}{2}H_{i+1} ]G
\Big\}\cr&&
-p_0\sum_{i=0}^\infty\theta^{i/2}ic_iH_iG .
\end{eqnarray}

\noindent Multiplying by $H_n(y)$ and integrating with respect to $y$ then gives (dividing by $\theta^{n/2}$ and formally
defining $c_{-1}:=0$)
\begin{eqnarray}
0 &=& -nc_n p_0 -\fracpp{x}\Big\{ \frac{1}{2}c_{n-1}g p_0
+\theta c_n f p_0 \nonumber\\&&
+ (n+1)\theta c_{n+1}g p_0
\Big\}.
\end{eqnarray}

\noindent As Eq.~(\ref{def_p0}) implies $\p_x (gp_0) = \frac{2f}{g}p_0$, we can get rid of the factor $p_0$. In terms
of the operator
\begin{eqnarray}
L &:=& \frac{g}{2}\frac{\p}{\p x}+\frac{f}{g}
\end{eqnarray}

\noindent this leads to
\begin{eqnarray}
0 = nc_n+L\Big[c_{n-1} 
+ \frac{2f}{g}c_n +2(n+1)c_{n+1}\Big].
\end{eqnarray}

\noindent For $n=0$, this equation does not provide any additional information, because it evaluates to
$0=L[(f/g)c_0+c_1]$, which, according to Eq.~(\ref{relation_c1_c0}), is fulfilled for all values of $\theta$.
Therefore, we only need to look at $n>0$ in the following. Inserting Eq.~(\ref{def_c_i}) and sorting terms by powers of
$\theta$ then yields
\begin{eqnarray}
0 &=& \Big\{ nc_{n,0}+Lc_{n-1,0} \Big\}\cr
  && +\sum_{i=1}^\infty \theta^i \Big\{
nc_{n,i}+L\big[c_{n-1,i} +\frac{2f}{g}c_{n,i-1}\cr
&&\qquad\qquad +2(n+1)c_{n+1,i-1}\big]
\Big\}.
\end{eqnarray}

\noindent As this equation must hold for arbitrary $\theta$, all expressions in curly brackets must vanish individually.
The first of these expressions, together with $c_{0,0}\equiv 1$, allows us to calculate all coefficients $c_{n,0}$,
\begin{eqnarray}
c_{n,0} &=& -\frac{1}{n} L c_{n-1,0} \;=\; \ldots \;=\; \frac{1}{n!}(-L)^n\cdot 1.
\end{eqnarray}

\noindent Explicitely one finds
\begin{subequations}
\begin{eqnarray}
c_{1,0} &=& -\frac{f}{g},\\
c_{2,0} &=& \frac{1}{2}\Big[\frac{g}{2}\Big(\frac{f}{g}\Big)'+\Big(\frac{f}{g}\Big)^2\Big],\\
&\vdots&.\nonumber
\end{eqnarray}
\end{subequations}

\noindent Similarly, all coefficients $c_{n,1}$ (and subsequently $c_{n,2}$, $c_{n,3}$, $\ldots$) can be calculated using
\begin{eqnarray}
c_{n,i} &=& -\frac{1}{n} L\big[c_{n-1,i} +\frac{2f}{g}c_{n,i-1}\cr
&&\qquad\qquad +2(n+1)c_{n+1,i-1}\big].
\end{eqnarray}

\noindent However, to start the iterative calculation of $c_{1,i}$, $c_{2,i}$, $\ldots$, we need the coefficient $c_{0,i}$,
which can be obtained as follows. For $n=1$, we may use Eq.~(\ref{relation_c1i_c0i}) to epress the left-hand side of the above
equation by $-(f/g)c_{0,i}$. This leads to
\begin{eqnarray}
\frac{\p}{\p x}c_{0,i} &=& -(\frac{\p}{\p x}+\frac{2f}{g^2})\big[\frac{2f}{g}c_{1,i-1} +4c_{2,i-1}\big].
\end{eqnarray}

\noindent We thus find
\begin{eqnarray}\label{def_c_0i}
c_{0,i} &=& C_i+c^*_{0,i}
\end{eqnarray}
\noindent with
\begin{eqnarray}
c^*_{0,i} &=& -\int_0^x(\frac{\p}{\p s}+\frac{2f(s)}{g^2(s)})\cr
&&\times\big[\frac{2f(s)}{g(s)}c_{1,i-1}(s) +4c_{2,i-1}(s)\big]\,ds,
\end{eqnarray}

\noindent where $C_i$ is an integration constant, which can be determined by using the fact that
$\int_xp_0c_{0,i}$ vanishes for $i>0$ (see Appendix~\ref{app_mean_cij}). Multiplying Eq.~(\ref{def_c_0i}) by $p_0$
and integrating with respect to $x$, therefore, yields
\begin{eqnarray}
C_i &=& -\int_{-\infty}^\infty p_0(x)c^*_{0,i}(x)\,dx.
\end{eqnarray}

\noindent To summarize results: We now have equations for all coefficients $c_{i,j}$ and for all integration constants
$C_i$. But, as noted above, these quantities need to be finite to ensure the validity of Eq.~(\ref{mean_Hn}).
We thus need to presume smooth and finite functions $g$ and $f/g$. Additionally, the limit density $p_0$ needs to decay
sufficiently fast, to ensure finite values $C_i$.

As a final remark: The result for the coefficient $c_{0,1}$, which is found to be
\begin{eqnarray}
c_{0,1} &=&  -g\left(\frac{f}{g}\right)'-\Big(\frac{f}{g}\Big)^2\nonumber\\
        &&   -\int_{-\infty\!}^\infty p_0(s) \Big[\frac{f(s)}{g(s)}\Big]^2\,ds,
\end{eqnarray}

\noindent may be checked for correctness using one of the small-$\theta$ approximations for $p(x)$ that are available in the
literature (see, e.g., \cite{haenggi95}). These approximations are known to correctly account for the first order
terms in $\theta$. Therefore, when expanding one of them into a power-series in $\theta$, the first order term needs
to equal $\theta p_0c_{0,1}$ --- which indeed is found to be the case.

\section{Integrals $\int_xp_0c_{i,j}$}
\label{app_mean_cij}

\noindent Integrating Eq.~(\ref{def_expansion}) with respect to $x$, inserting Eq.~(\ref{def_c_i}) and noting
$p(y)\equiv G(y)$ and $1\equiv H_0(y)$ leads to
\begin{eqnarray}
H_0(y) &=& \sum_{i=0}^\infty \theta^{i/2} H_i(y)\cr
&&\times \sum_{j=0}^\infty \theta^j\int_{-\infty}^\infty p_0(x)\, c_{i,j}(x)\,dx.
\end{eqnarray}

\noindent Since the functions $H_i$ are independent, it first follows
\begin{eqnarray}
1 &=& 1+\sum_{j=1}^\infty \theta^j\int_{-\infty}^\infty p_0(x)\, c_{0,j}(x)\,dx,\\
0 &=& \sum_{j=0}^\infty \theta^j\int_{-\infty}^\infty p_0(x)\, c_{i,j}(x)\,dx,\quad i>0,
\end{eqnarray}

\noindent where $c_{0,0}\equiv 1$, implying $\int_xp_0c_{0,0}=1$, has been used in the first equation. As these
equations must hold for arbitrary $\theta$, it further follows
\begin{eqnarray}
0 &=& \int_{-\infty}^\infty p_0(x)\, c_{0,j}(x)\,dx,\quad j>0,\\
0 &=& \int_{-\infty}^\infty p_0(x)\, c_{i,j}(x)\,dx,\quad i>0.
\end{eqnarray}

\section{Autocorrelation of $X(t)$}
\label{app_autocorr}

\noindent With the autocorrelation of $X(t)$ given by
\begin{eqnarray}
A(\tau) &:=& \left<X(\tau)X(0)\right>,
\end{eqnarray}

\noindent one first finds
\begin{eqnarray}
\Delta A(\tau) &:=& A(\tau)-A(0)\nonumber\\[.3em]
               &=& \left<[X(\tau)-X(0)]X(0)\right>\nonumber\\[.3em]
               &=& \int_{x,x'}(x'-x)x\, p(x',\tau;x,0)\,dx'\,dx.
\end{eqnarray}

\noindent Using $p(x',\tau;x,0)\!=\!p(x,0)p(x',\tau|x,0)$ then yields the connection to $M^{(1)}$,
\begin{eqnarray}
\Delta A(t) &=& \int_{x}p(x,0)\,x \int_{x'}  (x'-x)p(x',\tau|x,0)\,dx'\,dx\nonumber\\[.3em]
            &=& \int_{x}p(x,0)\,x M^{(1)}(\tau,x)\,dx.
\end{eqnarray}

\noindent With Eq.~(\ref{Mk_lambda_r_final}) one thus finds [up to order $O(\eps^3)$]
\begin{eqnarray}
\Delta A(\tau) &=& \sum_{i=1}^3  \left[\int_{x} x \lambda_i^{(1)}(x)\,p(x,0)\,dx\right]r_i(\tau)\nonumber\\[.3em]
               &=& \sum_{i=1}^3  \lambda_i r_i(\tau),
\end{eqnarray}

\noindent with unknown but constant coefficients $\lambda_i$.


\end{document}